\documentclass[]{aastex631}

\def\cof {$^{12}\mathrm{CO}~(J=1\rightarrow0)$}

\def\largestQ1{G034.4$+$01.0}

\def\msun{$M_{\odot}$}

\def\cofs {$^{12}\mathrm{CO}$}
\def\coss {$^{13}\mathrm{CO}$}
\def\cots{$\mathrm{C}^{18}\mathrm{O}$}
\def\deg  {\ifmmode {^\circ}\else {$^\circ$}\fi}

\def\kms     {km~s$^{-1}$}

\usepackage{booktabs}
\usepackage{chngcntr}

\date{\today}
\submitjournal{ApJ}

\shorttitle{Molecular Cloud Samples}
\shortauthors{Yan et al.} 

\begin{document}



\title{Dependence of Molecular Cloud Samples on Angular Resolution, Sensitivity, and Algorithms}

\correspondingauthor{Ji Yang}
\email{jiyang@pmo.ac.cn,qzyan@pmo.ac.cn}

\author[0000-0003-4586-7751]{Qing-Zeng Yan}
\altaffiliation{Chang Yu-Che Fellow of Purple Mountain Observatory}
\affil{Purple Mountain Observatory and Key Laboratory of Radio Astronomy,\\
 Chinese Academy of Sciences, 10 Yuanhua Road, Qixia District, Nanjing 210033, People's Republic of China}

\author[0000-0001-7768-7320]{Ji Yang}
\affil{Purple Mountain Observatory and Key Laboratory of Radio Astronomy,\\
 Chinese Academy of Sciences, 10 Yuanhua Road, Qixia District, Nanjing 210033, People's Republic of China}

 \author[0000-0002-0197-470X]{Yang Su }
 \affil{Purple Mountain Observatory and Key Laboratory of Radio Astronomy,\\
 Chinese Academy of Sciences, 10 Yuanhua Road, Qixia District, Nanjing 210033, People's Republic of China}

 \author[0000-0002-3904-1622]{Yan Sun}
\affil{Purple Mountain Observatory and Key Laboratory of Radio Astronomy,\\
 Chinese Academy of Sciences, 10 Yuanhua Road, Qixia District, Nanjing 210033, People's Republic of China}

 \author[0000-0003-2418-3350]{Xin Zhou}
\affil{Purple Mountain Observatory and Key Laboratory of Radio Astronomy,\\
 Chinese Academy of Sciences, 10 Yuanhua Road, Qixia District, Nanjing 210033, People's Republic of China}
 
\author[0000-0001-5602-3306]{Ye Xu}
\affil{Purple Mountain Observatory and Key Laboratory of Radio Astronomy,\\
 Chinese Academy of Sciences, 10 Yuanhua Road, Qixia District, Nanjing 210033, People's Republic of China}

 \author[0000-0003-0746-7968]{Hongchi Wang}
\affil{Purple Mountain Observatory and Key Laboratory of Radio Astronomy,\\
 Chinese Academy of Sciences, 10 Yuanhua Road, Qixia District, Nanjing 210033, People's Republic of China}

 \author[0000-0003-2549-7247]{Shaobo Zhang}
\affil{Purple Mountain Observatory and Key Laboratory of Radio Astronomy,\\
 Chinese Academy of Sciences, 10 Yuanhua Road, Qixia District, Nanjing 210033, People's Republic of China}

\author[0000-0003-0849-0692]{Zhiwei Chen}
\affil{Purple Mountain Observatory and Key Laboratory of Radio Astronomy,\\
 Chinese Academy of Sciences, 10 Yuanhua Road, Qixia District, Nanjing 210033, People's Republic of China}






\begin{abstract}  
In this work, we investigate the observational and algorithmic effects on molecular cloud samples identified from position-position-velocity (PPV) space. By smoothing and cutting off the high quality data of the Milky Way Imaging Scroll Painting (MWISP) survey,  we extract various molecular cloud samples from those altered data with the DBSCAN (density-based spatial clustering of applications with noise) algorithm. Those molecular cloud samples are subsequently used to gauge the significance of sensitivity, angular/velocity resolution, and DBSCAN parameters. Two additional surveys, the FCRAO Outer Galaxy Survey (OGS) and the CfA-Chile 1.2 m complete CO (CfA-Chile) survey, are used to verify the MWISP results. We found that molecular cloud catalogs are not unique and the boundary and therefore the number shows strong variation with angular resolution and sensitivity. At low angular resolution (large beam sizes), molecular clouds merge together in PPV space, while low sensitivity (high cutoffs) misses small faint molecular clouds and takes bright parts of large molecular clouds as single  ones.  At high angular resolution and sensitivity, giant molecular clouds (GMCs) are resolved into individual clouds, and their diffuse components are also revealed. Consequently, GMCs are more appropriately interpreted as clusters or aggregates of molecular clouds, i.e., GMCs represent molecular cloud samples themselves. 
\end{abstract}










\section{Introduction} \label{sec:intro}

Molecular clouds \citep[see][for a review]{2015ARA&A..53..583H} are usually observed in three-dimensional position-position-velocity (PPV) space or two-dimensional dust continuum \citep[e.g.,][]{ 2007A&A...471..103B,2015A&A...584A..91K,2019A&A...621A..42A}. PPV data provide valuable information about properties of molecular clouds and activities in them, such as densities \citep[e.g.,][]{2017A&A...605L...5K,2021ApJ...920..126E}, temperatures \citep[e.g.,][]{2019ApJS..243...25W,2021A&A...650A.164M}, and molecular outflows \citep[e.g.,][]{1980ApJ...239L..17S,2001ApJ...552L.167Z,2018ApJS..235....3Y,2020ApJS..248...15Z,2022A&A...658A.160Y}, as well as large-scale structures of the Milky Way \citep[e.g.,][]{2001ApJ...547..792D,2015ApJ...798L..27S,2017ApJS..229...24D}. Missing one dimension of position, PPV data are incomplete compared with position-position-position (PPP) data and are unable to distinguish molecular clouds in crowded regions, such as the first Galactic quadrant, and velocity degenerated directions, such as the Galactic anti-center.  In addition to this dimensional incompleteness, PPV observations are also subjected to  effects due to angular/velocity resolution and sensitivity, and observational effects on molecular cloud studies have been brought into focus recently \citep{2021ApJ...910..109Y,2021A&A...653A.157L,2021arXiv211108125P}.











 Shortly after the first detection of CO emission \citep{1970ApJ...161L..43W}, molecular clouds are preliminarily studied by observing prominent star-forming regions or by coarse imaging of the Galactic plane. Those researches included both low-mass star-forming regions, such as  Perseus OB2 \citep{1977PASJ...29...53K}, the Taurus molecular cloud complex \citep[e.g.,][]{1985ApJ...298..818M,1987ApJS...63..645U,2008ApJ...680..428G}, and  L1287 \citep{1992ApJ...386..618Y}, and high-mass star-forming regions, such as the Orion Nebula  \citep[e.g., ][]{1974ApJ...190..557L,1977ApJ...215..521K}, Cynus X \citep{1977PhDT.......123C,1992ApJS...81..267L},  Cepheus A \citep{1980ApJ...240L.149R}, and  Monoceros \citep{1975ApJ...199...79K,1996A&A...315..578O,2005A&A...430..523W}.

At this stage, molecular cloud samples were mainly built with empirical methods. For instance, to study turbulence in multiple scales, \citet{1981MNRAS.194..809L} referred to molecular clouds as specific regions, which could be cloud complexes, individual molecular clouds in cloud complexes, or even clumps in individual molecular clouds. Alternatively, \citet{1987ApJ...319..730S} defined molecular clouds as objects within closed surfaces in PPV space. They identified 273 molecular clouds in the first Galactic quadrant, and due to the 3\arcmin\ survey spacing, there was no need to refine the definition of molecular clouds.

In order to obtain a comprehensive understanding of molecular clouds in the Milky Way, many large-scale CO surveys have been conducted. A list of CO surveys is compiled by \citet{2015ARA&A..53..583H}, and among them, the CfA-Chile 1.2 m complete CO (CfA-Chile) survey \citep{2001ApJ...547..792D} possesses the largest sky coverage. Other surveys, such as the Five College Radio Astronomy Observatory Outer Galaxy Survey \citep[OGS,][]{1998ApJS..115..241H} and the Boston University-Five College Radio Astronomy Observatory Galactic Ring Survey \citep[GRS,][]{2006ApJS..163..145J}, reveal many more details of molecular clouds. In the last two decades, CO maps of the Galactic plane have been constantly refined by newly conducted surveys, such as the SEDIGISM \citep[structure, excitation, and dynamics of the inner Galactic interstellar medium,][]{2017A&A...601A.124S} survey and the Milky Way Imaging Scroll Painting \citep[MWISP,][]{2019ApJS..240....9S} survey.

With the progress of large-scale CO surveys and the development of sophisticated algorithms, high quality molecular cloud samples are now achievable. For instance, \citet{2016ApJ...822...52R} provided a uniform catalog of 1064 with the CfA-Chile survey. Using the same \cofs\ data but a different algorithm, \citet {2017ApJ...834...57M} identified 8107 molecular clouds in the Galactic plane. However, this number is smaller than that of the \cofs\ catalog \citep[14592,][]{2003ApJS..144...47B} extracted from the OGS in the second Galactic quadrant. Using \cofs\ maps of the MWISP survey, \citet{2020ApJ...898...80Y} identified about 4,000 local molecular clouds using the DBSCAN (density-based spatial clustering of applications with noise)  algorithm in the first Galactic quadrant. As to other molecular lines, \citet{2019MNRAS.483.4291C} identified more than 85000 clouds using the JCMT $^{12}\mathrm{CO}~(J=3\rightarrow2)$ High-Resolution Survey \citep[COHRS,][]{2013ApJS..209....8D} and the Spectral Clustering for Interstellar Molecular Emission Segmentation \citep[SCIMES,][]{2015MNRAS.454.2067C} algorithm. Recently, \citet{2021MNRAS.500.3027D} compiled a catalog of 10,663 $^{13}\mathrm{CO}~(J=2\rightarrow1)$ clouds with the SEDIGISM survey.

Source finding in PPV space, however, can be significantly affected by sensitivity, resolution, and algorithms. First of all, a general cutoff is  imposed on PPV data cubes, and evidently, sensitivity determines the fraction of flux observed \citep{2021ApJ...910..109Y}. For example, \citet{2012PASA...29..276W} tested the \texttt{DUCHAMP} source finder and found that \texttt{DUCHAMP} is useful in locating the source in PPV space but properties of those sources are systematically affected by the noise. \citet{2009ApJ...699L.134P} examined the robustness of \texttt{Clumpfind} \citep{1994ApJ...428..693W} in deriving the mass spectrum of molecular clumps, concluding that \texttt{Clumpfind} is not suitable for deriving physically meaningful mass functions and that the \coss\ mass spectrum is significantly affected by the observation resolution. \citet{2020ApJ...898...80Y} compared molecular cloud samples with three algorithms: DBSCAN algorithm \citep{citeDBSCAN,2020ApJ...898...80Y}, hierarchical DBSCAN \citep[HDBSCAN,][]{citeHDBSCAN}, and SCIMES \citep{2015MNRAS.454.2067C}, and the results are significantly different. Usually, algorithms for building molecular cloud catalogs are  designed to keep the balance between noise contamination and flux completeness. They found that the definition of molecular clouds with HDBSCAN is non-uniform and SCIMES misses a significant fraction of flux in the process of decomposing dendrogram trunks. Consequently, they recommended using DBSCAN to identify molecular clouds from MWISP data cubes. 

 \citet{2021ApJ...910..109Y} examined the angular resolution and sensitivity effects on two fundamental properties of molecular clouds, the beam filling factor and the flux completeness. By comparing molecular cloud samples extracted from observations at various angular resolution and sensitivity, they found that the beam size is directly related to the brightness of molecular clouds, and the signal-to-noise ratio determines the flux completeness. Quantitative analyses indicate that the relationship between angular resolution and the beam filling factor is similar to that between sensitivity and the flux completeness.


In molecular cloud catalogs, giant molecular clouds (GMCs), one molecular cloud population with the largest size and mass \citep[see][for a recent review]{2020SSRv..216...50C}, have been investigated heavily. GMCs are initially studied in the Milky Way \citep[e.g.,][]{1979IAUS...84...35S,2016ApJ...822...52R,2020ApJ...898....3L}, and due to the unprecedented resolution and sensitivity of the Atacama Large Millimeter/submillimeter Array (ALMA), GMC investigations are now feasible in the Local Group galaxies \citep{2015ApJ...808...99R,2018ApJ...857...19F,2021MNRAS.502.1218R}. For instance, \citet{2021MNRAS.502.1218R} reviewed the extragalactic GMC catalogs (see references therein) and identified 4986 GMCs based on the Physics at High Angular resolution in Nearby Galaxies (PHANGS) project performed with ALMA. They found that physical properties of GMCs, such as the surface density, line width, and internal pressure, vary among galaxies, consistent with previous studies \citep[e.g.,][]{2013ApJ...779...46H}.

Mapping from PPV sources to  PPP sources is not one-to-one, and we may never obtain authentic molecular cloud samples based on PPV space. For example, \citet{2016MNRAS.458.3667D} found that the most important factor that affects PPV data is the angular/spatial resolution, and clouds in PPV space may suffer from strong blending. This blending prevents precise matches between PPV and PPP molecular clouds \citep{2017MNRAS.470.4261D}, as well as degrading statistical results regarding molecular cloud properties \citep{2016MNRAS.458.2443P}.


In this work, we further explore factors that affect molecular cloud samples drawn from PPV space, including the angular/velocity resolution, sensitivity, and algorithms. Starting from the simplest cases, we alter the MWISP data by smoothing and clipping, corresponding to large beam sizes, low velocity resolution, and low sensitivity, and use the DBSCAN algorithm to draw molecular cloud samples from those altered MWISP data. Two other surveys, the OGS and the CfA-Chile survey, are employed for the verification of MWISP results. Cloud samples derived with two alternative algorithms, HDBSCAN and SCIMES, are also used for comparison.





This paper is organized as follows. Section \ref{sec:data} describes the CO data and the cloud identification method. Section \ref{sec:result} presents results of molecular cloud samples based on altered data of the MWISP survey, as well as comparisons with the OGS and the CfA-Chile survey.  Section \ref{sec:gmc} displays results of three GMCs, W3, W5, and NGC 7538, in three CO surveys. Discussions are presented in Section \ref{sec:discuss}, and we summarize  conclusions in Section \ref{sec:summary}.







\section{CO Data and Molecular Cloud Identification} 
\label{sec:data}

\begin{deluxetable}{cccccccc}
\tablecaption{Observational parameters of the three \cof\ surveys. \label{Tab:threeSurvey}}
\tablehead{
\colhead{Survey} & \colhead{$l$\tablenotemark{a}} & \colhead{$b$\tablenotemark{a}}  &  \colhead{$V_{\rm LSR}$\tablenotemark{a}}   & \colhead{Beam Size} & \colhead{Rms Noise} & 
\colhead{Velocity Resolution} & \colhead{Reference} \\
\colhead{   } & \colhead{(\deg)} & \colhead{(\deg)} & (\kms) &  & (K)   &  (\kms) &
}

\startdata
CfA-Chile  &   [110, 141.54]   &  [-2, 4]   &   [-137, 32] & $8'$   & 0.1   &  1.3 &  \citet{2001ApJ...547..792D}  \\ 
 OGS  &  [110, 141.54]  & [-2, 4] &  [-137, 32]  &  $45''$   & 0.6   &   0.98 &   \citet{1998ApJS..115..241H} \\ 
MWISP  &   [110, 141.54] &  [-2, 4]  & [-137, 32] &  $49''$  &   0.5  & 0.167 &  \citet{2019ApJS..240....9S}  \\ 
\enddata
\tablenotetext{a}{We select a region that is commonly covered by all three surveys.}
\end{deluxetable}

\subsection{CO data} 
 \label{sec:codata}

The CO data are part of the MWISP CO survey \citep{2019ApJS..240....9S}. The MWISP CO survey records three CO isotopologue lines, and we use the most extended \cof\ emission to identify molecular cloud samples. The angular resolution of \cofs\ is about 50\arcsec, at an rms noise level of about 0.5 K. The large-scale sky coverage and high sensitivity of the MWISP CO survey allow us to construct observations with low angular resolution and low  sensitivity through smoothing and clipping, respectively.

In total, we produce three sets of observations, and each set corresponds to one variation of spatial resolution, velocity resolution, or  sensitivity. The beam size and velocity resolution vary by factors from 2 to 10 with a step of 1. As to the sensitivity, we use the cutoff as a proxy, from 3$\sigma$ to 11$\sigma$ with a step of 1$\sigma$, where $\sigma$ is the rms noise. In total, we reproduced 27 cases of observations.

For the convenience of comparisons, we chose a region that has been commonly mapped by the MWISP survey, the CfA-Chile survey, and the OGS. The $l$, $b$, and $V_{\rm LSR}$ ranges are [110\deg, 141\fdg54], [-2\deg, 4\deg] and [-137, 32] \kms, respectively. In addition, the velocity crowding in this region of the second Galactic quadrant is less severe. See Table \ref{Tab:threeSurvey} for a summary of observational parameters of these three surveys. The MWISP survey and the OGS have close angular resolution ($\sim$$50''$) and rms noises ($\sim$0.5 K), but the velocity resolution of the MWISP survey is much finer, yielding a higher sensitivity. The beam size of the CfA-Chile survey is about ten times larger compared with the MWISP survey and the OGS, but its sensitivity is much higher.


The publicly available OGS data is in the unit of $T_{\rm R}^*$, obtained by adding corrections of the forward spillover and scattering efficiency of the telescope \citep{1981ApJ...250..341K} to the antenna temperature $T_{\rm A}^*$. For the OGS data,  $T_{\rm R}^* = T_{\rm A}^*/0.7$, and considering the main beam efficiency (0.45) of the FCRAO 14 m antenna,  the main beam brightness temperature, $T_{\rm mb}$, equals $1.556T_{\rm R}^*$. In other words, a factor of 1.556 is used to convert the OGS data to the main beam brightness temperature \citep[see also][]{2021ApJS..256...32S}.





\subsection{Molecular Cloud Identification}
 \label{sec:cloudIdentification}

The algorithm used to extract molecular cloud samples is DBSCAN\footnote{\href {https://scikit-learn.org/stable/modules/generated/sklearn.cluster.DBSCAN.html}{https://scikit-learn.org/stable/modules/generated/sklearn.cluster.DBSCAN.html}} \citep{2020ApJ...898...80Y}. DBSCAN identifies independent structures in PPV space based on connectivity of voxels, ignoring internal structures, such as cores or clumps. We believe this definition of molecular clouds is more appropriate, and without additional information, such as distances, connected structures in PPV space should be recognized as single objects. Consequently, DBSCAN molecular clouds are used to examine the variation of molecular cloud samples in this work.


The two parameters of DBSCAN are referred to as MinPts and connectivity when applied to PPV data cubes. Connectivity defines the linkage between two points, while MinPts defines core points. Two points are said to be linked if the distance between them is less than or equal to a threshold specified by connectivity, and for data cubes, there are three reasonable distance thresholds, which are 1, $\sqrt{2}$, or $\sqrt{3}$. A point is defined to be a core if the number of its linked points (including itself) is no less than the MinPts value. Connected core points together with their linked points form molecular clouds. Large connectivity distances require larger MinPts values to avoid including too many noises, but exceedingly large MinPts values make DBSCAN only pick  up compact bright regions, missing a large fraction of flux.

Due to its highest compactness, we use the recommended parameter combination of connectivity 1 and MinPts 4. Two other parameter combinations are also examined to see effects of DBSCAN parameters. Before performing DBSCAN, all points with brightness temperatures below  a cutoff are removed. And four extra post criteria are applied to filter noise DBSCAN structures, as suggested by \citet{2020ApJ...898...80Y}. Those criteria are (1) the voxel number of a molecular cloud is at least 16, (2) the peak brightness temperature of a molecular cloud is at least 5$\sigma$, (3) the integrated map of a molecular cloud contains a compact 2$\times$2 (in pixels) regions, and (4) the velocity channel number is at least 3.

The DBSCAN algorithm is applied to the raw data cube and 27 altered ones (see Section \ref{sec:codata}) of the MWISP survey, as well as data cubes of the CfA-Chile survey and the OGS, to reveal the effect of the angular resolution and sensitivity.

 \begin{figure}[ht!]
 \plotone{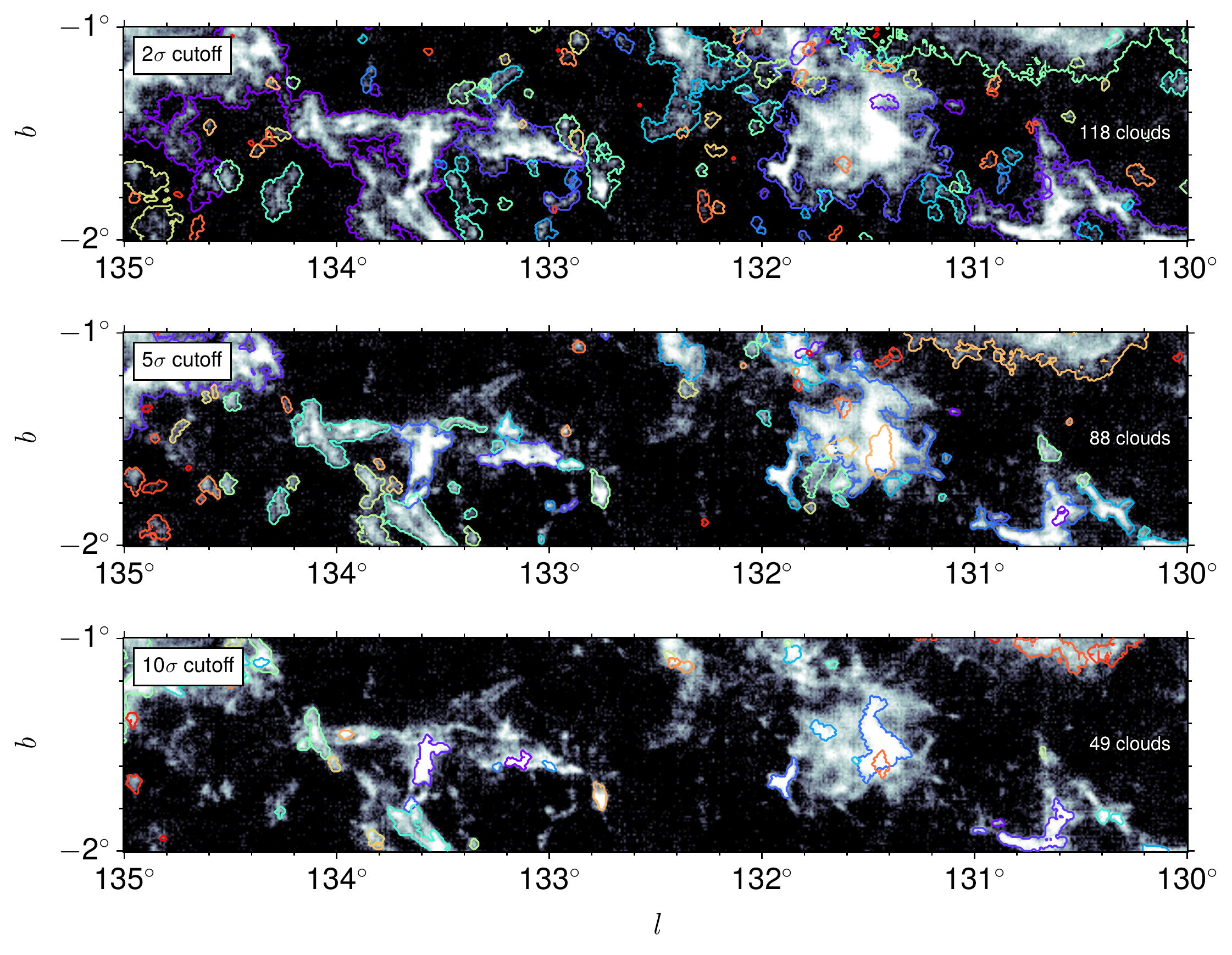}
\caption{Molecular clouds under different cutoffs of the MWISP survey. The number of molecular clouds only takes into account of those that have parts in the PPV range of  $130\deg \leq l  \leq 135\deg$, $-2\deg \leq b \leq-1\deg$, and $-20 \leq  V_{\rm LSR} \leq -10$ \kms. In the displayed three cases, from top to bottom,  118, 88, and 49 molecular clouds are identified, respectively. \label{fig:sensitivityImage}} 
\end{figure}

\section{Results}
\label{sec:result}



This section displays the variation of molecular cloud samples with respect to observations. To measure this variation, we examine the edge and number of molecular clouds, as well as their distributions of angular area and flux. In order to demonstrate the variation of molecular cloud edges clearly, we zoom in on the images,  showing a subregion of $130\deg \leq l  \leq 135\deg$, $-2\deg \leq b \leq-1\deg$, and $-20 \leq  V_{\rm LSR} \leq  -10$ \kms. However, plots of distributions use all samples in the whole region (see Table \ref{Tab:threeSurvey}).

We first, using the MWISP survey data, examine effects rooted in sensitivity, angular and velocity resolution, and the DBSCAN parameter settings. Secondly, as a verification of MWISP results, molecular cloud samples from the OGS and the CfA-Chile survey data are used for a comparison.


 \begin{figure}[ht!]
 \plotone{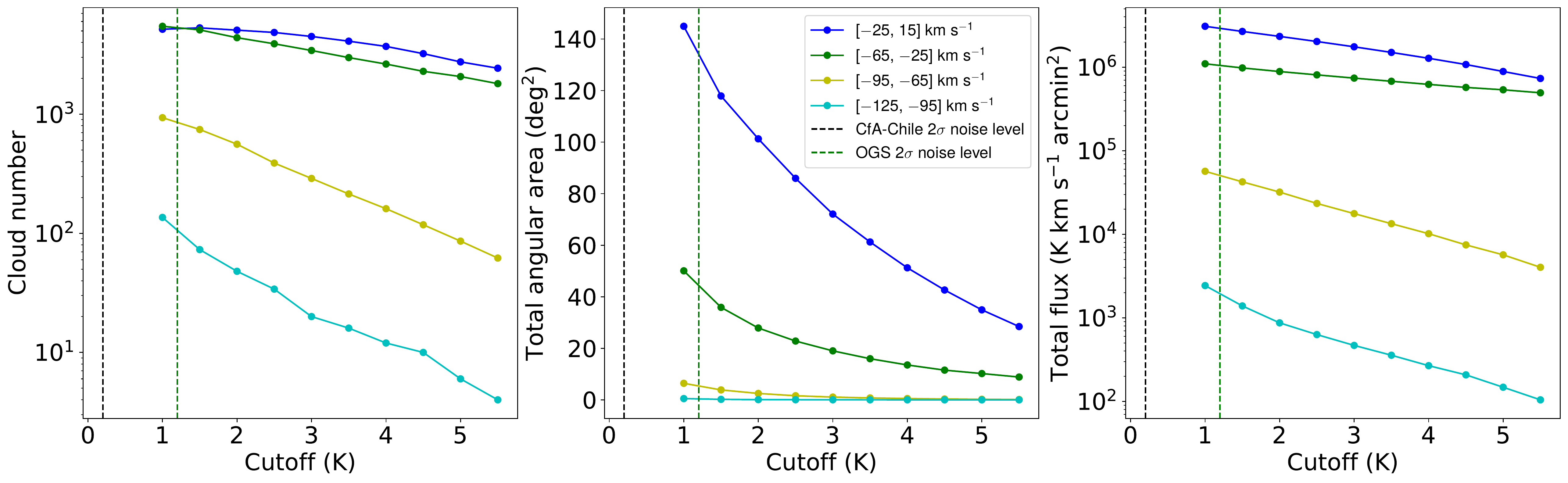}
\caption{Variation of molecular cloud number, the total angular area, and the total flux with respect to cutoffs. In order to see the distance effects, molecular cloud samples are split into four parts, approximately corresponding to four arm segments \citep{2021ApJS..256...32S}, the Local (-25 to 15 \kms), the Perseus ($-65$ to $-25$ \kms), the Outer ($-95$ to $-65$ \kms), and the Outer Scutum Centaurus (OSC, $-125$ to $-95$ \kms) arms. \label{fig:sensitivityPlot}} 
\end{figure}

\subsection{Sensitivity Effects}

We use the cutoff as a proxy of sensitivity. Low cutoffs correspond to high sensitivity, and conversely, high cutoffs correspond to low sensitivity. Figure \ref{fig:sensitivityImage} demonstrates that with the decrease of sensitivity, edges of molecular clouds shrink, many small molecular clouds are missed, and large molecular clouds are split into many smaller ones. 

\begin{figure}[ht!]
 \plotone{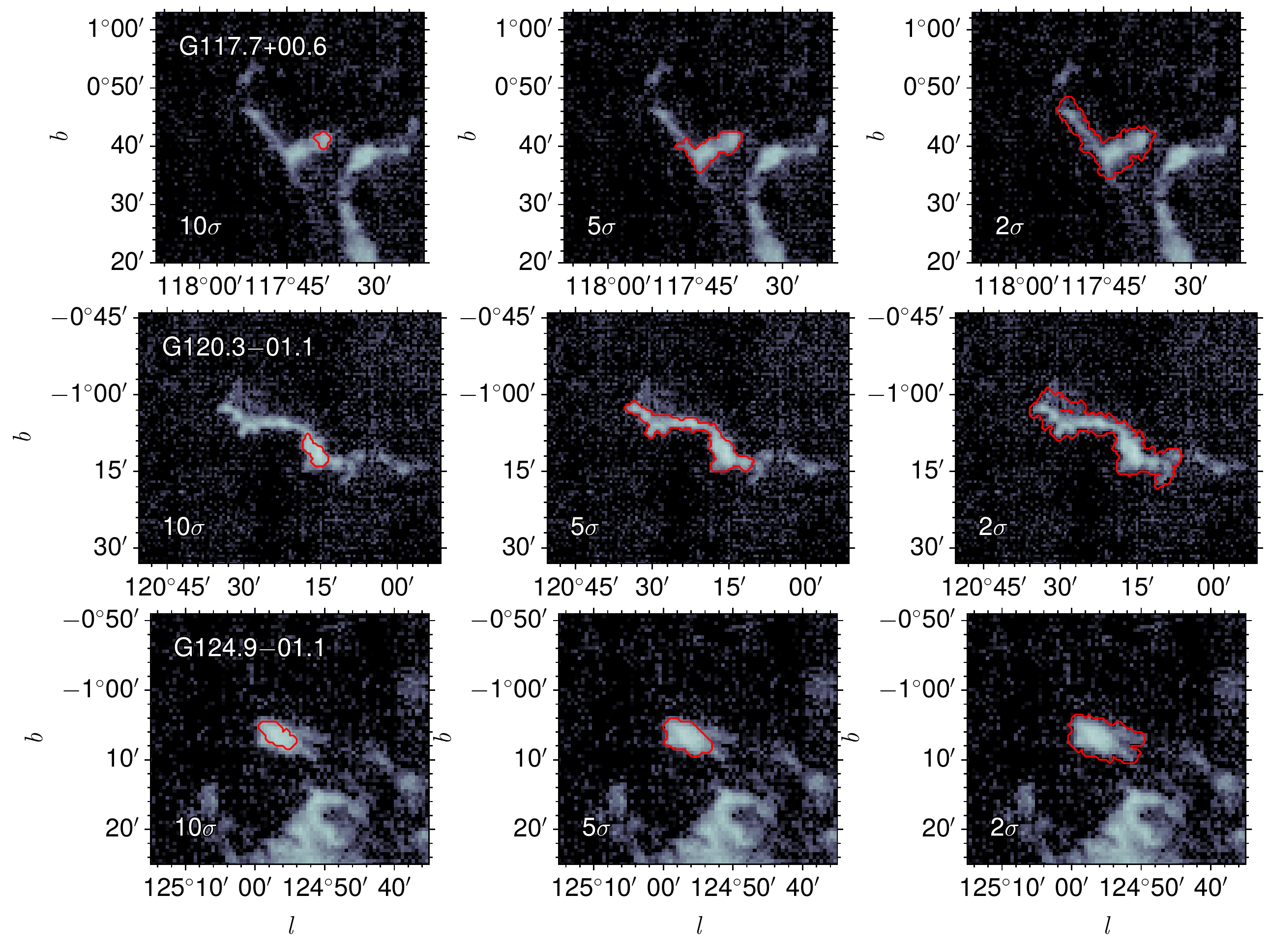}
\caption{The area variation of three molecular clouds, G117.7$+$00.6, G120.3$-$01.1, and G124.9$-$01.1, against the cutoffs. Three cases, 10$\sigma$,  5$\sigma$, and 2$\sigma$, are shown for each molecular cloud, and the background images are the integrated maps in the MWISP survey.  \label{fig:sensitivityAreaSingle}} 
\end{figure}


The variation of molecular cloud numbers against the cutoff is displayed in Figure \ref{fig:sensitivityPlot}. In order to demonstrate the distance effect, we divide molecular cloud samples into four groups, roughly corresponding to four arm segments \citep{2021ApJS..256...32S}: the Local (-25 to 15 \kms), the Perseus ($-65$ to $-25$ \kms), the Outer ($-95$ to $-65$ \kms),  and the Outer Scutum Centaurus (OSC, $-125$ to $-95$ \kms) arms. 

The definition of arms is empirical and inaccurate, but it has little effects on conclusions. Given the high sensitivity and large sky coverage, it is possible to determine CO arms more precisely with the MWISP survey. However, this is beyond the scope of this work and will be done in a future study.

Evidently, both far distances and low sensitivity decrease the number of molecular clouds. Interestingly, the number of local molecular clouds at 2$\sigma$ cutoff is slightly smaller than that at 3$\sigma$ cutoff. This convergence feature \citep{2021A&A...653A.157L} indicates that for local molecular clouds, the flux completeness is probably sufficiently high in the MWISP survey.

Moreover, sensitivity also affects the angular area significantly. Figure \ref{fig:sensitivityPlot} demonstrates the variation of the total angular area of molecular clouds with respect to the cutoff. Evidently, the total angular area decreases fast, particularly for the molecular clouds in the Local and Perseus arms.

 This effect of area decreasing can also been seen in molecular cloud individuals. Figure \ref{fig:sensitivityAreaSingle} shows the angular area of three molecular clouds at different sensitivity, and the angular area at the 2$\sigma$ cutoff is about 10 times of that at the 10$\sigma$ cutoff. To avoid velocity contamination, the displayed three molecular clouds are required to be unique ones along their sightlines.

\subsection{Angular Resolution}
\label{sec:ang}

Angular resolution has two effects on radio spectroscopy observations. First, low angular resolution yields low beam filling factors \citep{2021ApJ...910..109Y}, diminishing the brightness of molecular clouds. As a consequence, in order to collect the same amount of flux, high sensitivity is required for low angular resolution observations. Secondly, low angular resolution merges molecular clouds that are close in PPV space, degrading the quality of molecular cloud samples.



\begin{figure}[ht!]
 \plotone{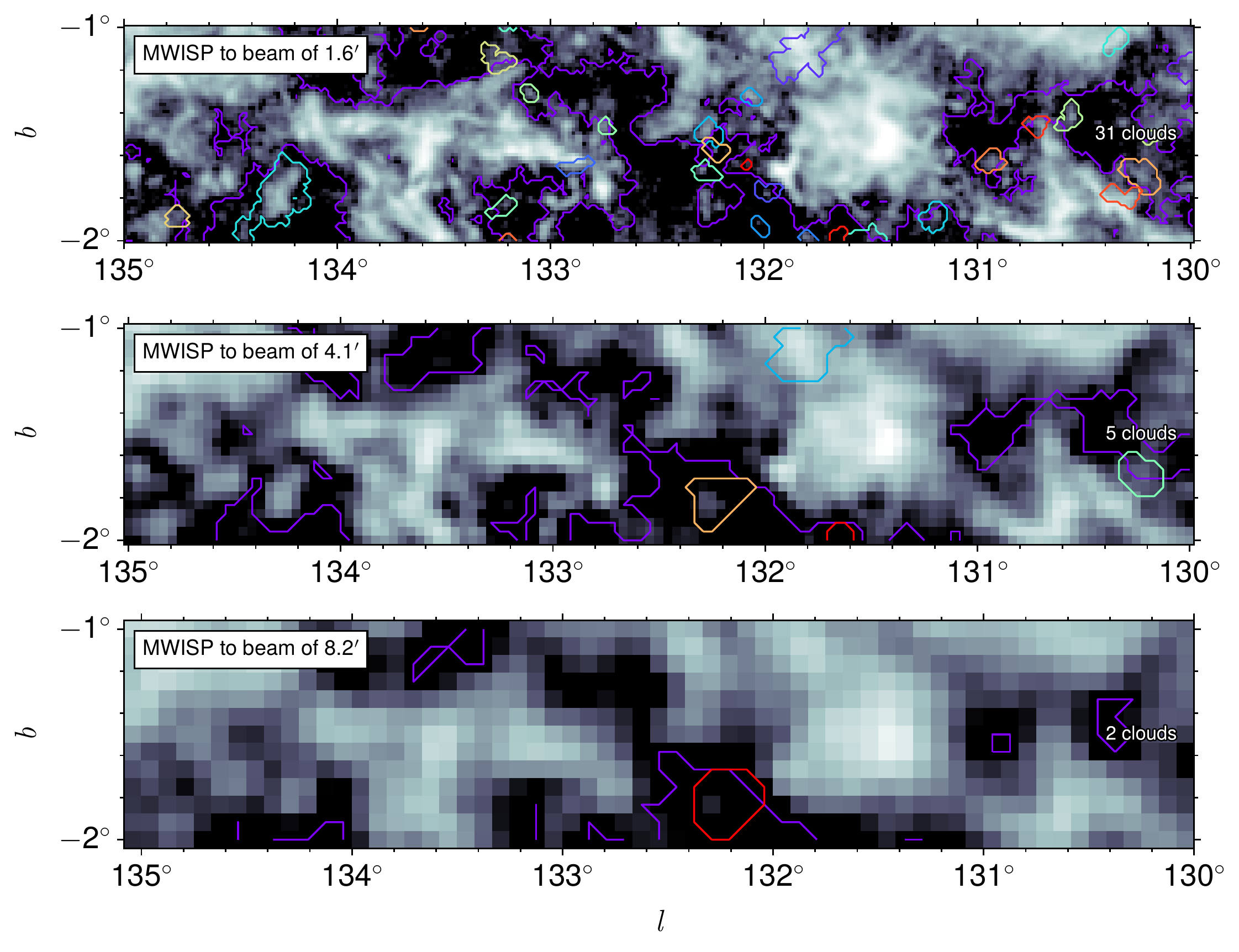}
\caption{Molecular clouds against spatial smoothing of the MWISP survey. The number of molecular clouds only takes into account of those that have parts in the PPV range of  $130\deg \leq l  \leq 135\deg$, $-2\deg \leq b \leq-1\deg$, and $-20 \leq  V_{\rm LSR} \leq -10$ \kms. In the displayed three cases, from top to bottom, 31, 5, and 2 molecular clouds are identified, respectively. \label{fig:smMWISP}} 
\end{figure}

Figure \ref{fig:smMWISP} displays images and samples of molecular clouds, smoothed to beam sizes of 1.6\arcmin, 4.1\arcmin, and 8.2\arcmin.  The number of molecular clouds decreases significantly. At 8.2\arcmin, which is close to the angular resolution of the CfA-Chile survey, only two molecular clouds were found, consistent with the results of the CfA-Chile survey, which yields only one single molecular cloud (see Section \ref{sec:threeco}).

\begin{figure}[ht!]
 \plotone{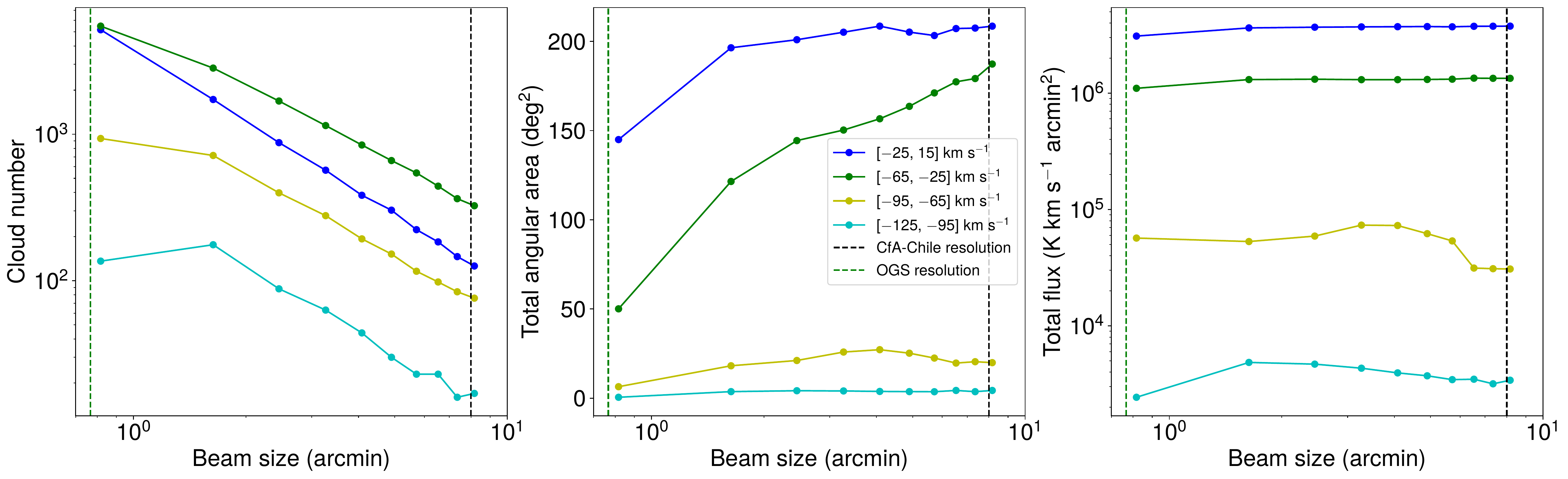}
\caption{Variation of molecular cloud number, the total angular area, and the total flux with respect to the beam size. In order to see the distance effects, molecular cloud samples are split into four parts, approximately corresponding to four arm segments \citep{2021ApJS..256...32S}, the Local (-25 to 15 \kms), the Perseus ($-65$ to $-25$ \kms), the Outer ($-95$ to $-65$ \kms), and the Outer Scutum Centaurus (OSC, $-125$ to $-95$ \kms) arms. \label{fig:smMWISPplot}} 
\end{figure}

The number of molecular clouds is roughly proportional to a power-law form of the beam size. As demonstrated in Figure \ref{fig:smMWISPplot}, in logarithmic form, the number of molecular clouds is linearly correlated with beam size. Interestingly, this pattern is similar for all four arm segments.


\subsection{Velocity Resolution}

In addition to the angular resolution, we also examine the dependence of molecular cloud samples on velocity resolution. Velocity resolution only affects one axis, and as expected, its effects on the PPV data are less significant compared with the angular resolution.

\begin{figure}[ht!]
 \plotone{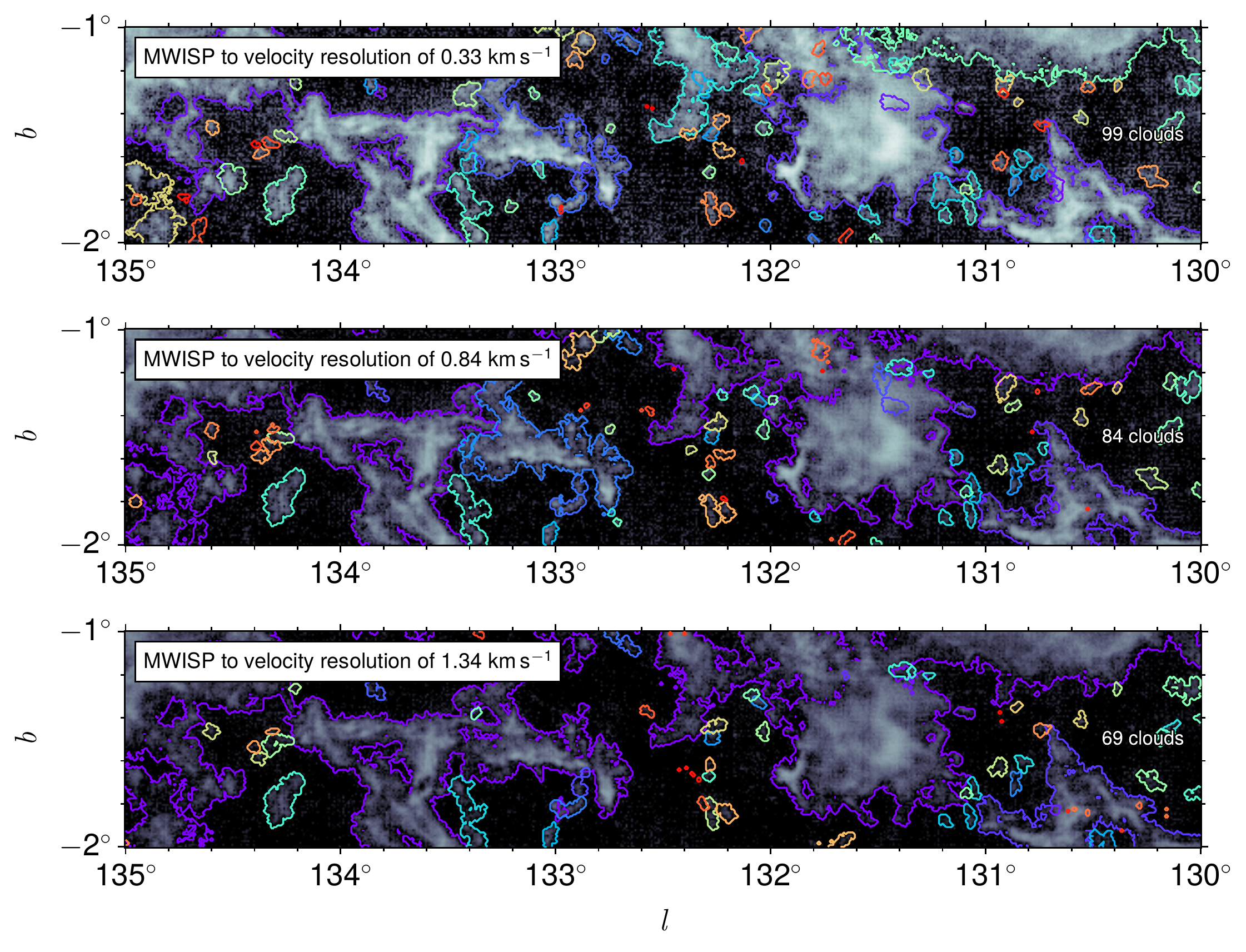}
\caption{ Molecular clouds of the MWISP survey against the velocity resolution. The number of molecular clouds only takes into account of those that have parts in the PPV range of  $130\deg \leq l \leq 135\deg$, $-2\deg \leq b \leq-1\deg$, and $-20 \leq  V_{\rm LSR} \leq -10$ \kms. In the displayed three cases, 0.33, 0.84, or 1.34 \kms, 99, 84, and 69 molecular clouds are identified, respectively. \label{fig:velsmImage} } 
\end{figure}

\begin{figure}[ht!]
 \plotone{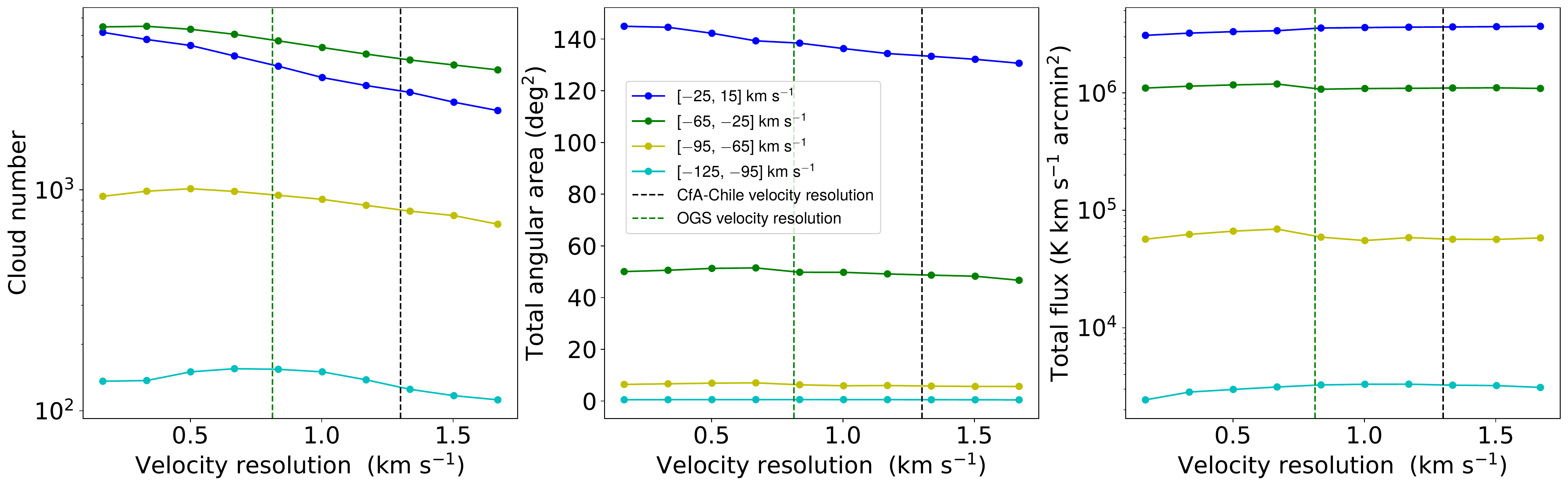}
\caption{Variation of molecular cloud number, the total angular area, and the total flux with respect to velocity resolution. In order to see the distance effects, molecular cloud samples are split into four parts, approximately corresponding to four arm segments \citep{2021ApJS..256...32S}, the Local (-25 to 15 \kms), the Perseus ($-65$ to $-25$ \kms), the Outer ($-95$ to $-65$ \kms), and the Outer Scutum Centaurus (OSC, $-125$ to $-95$ \kms) arms. \label{fig:velsmPlot}} 
\end{figure}

 Figure \ref{fig:velsmImage} shows images of three cases of smoothing along the velocity axis, resulting velocity resolution of 0.33, 0.84, or 1.34 \kms, respectively. The number of molecular clouds decreases slightly toward low velocity resolution, and some large molecular clouds merge into single ones. As shown in Figure \ref{fig:velsmPlot}, the trend of decreasing is roughly linear.



\subsection{DBSCAN parameter settings}

\citet{2020ApJ...898...80Y} compared molecular cloud samples identified with three DBSCAN parameter combinations in the first Galactic quadrant. Each combination corresponds to one connectivity and an appropriate MinPts, and the MinPts value is chosen to keep as much as flux and reject as many as noises.

\begin{figure}[ht!]
\plotone{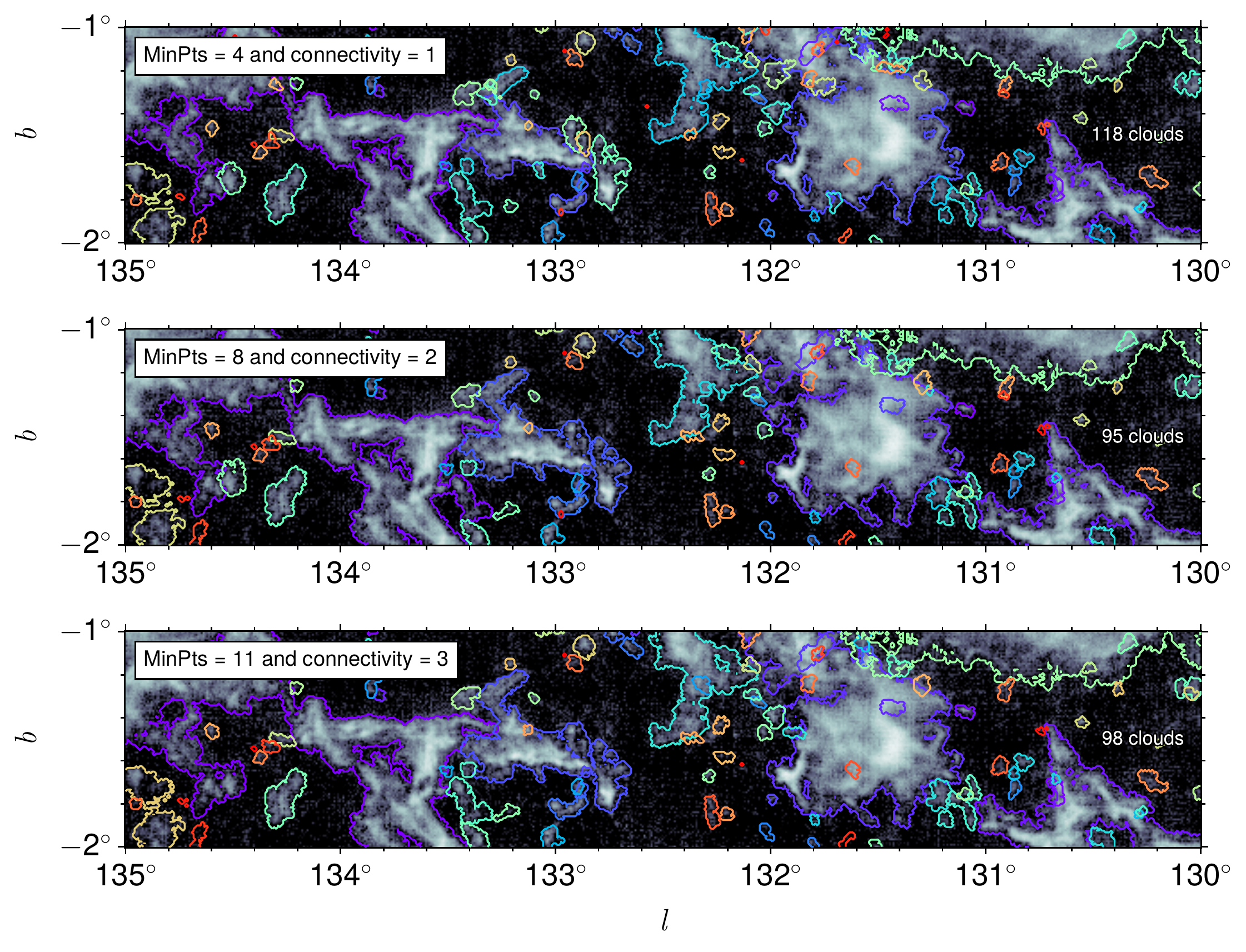}
\caption{Molecular clouds with three DBSCAN parameter combinations based on the MWISP survey. Each combination corresponds to one connectivity and an appropriate MinPts value. Connectivity 1 identifies slightly more molecular clouds than the rest two cases. \label{fig:dbscanpara}  } 
\end{figure}

To see the robustness of DBSCAN, we compare molecular clouds at different parameter combinations in the second Galactic quadrant. As demonstrated in Figure \ref{fig:dbscanpara}, differences between the results are insignificant. Molecular cloud samples with connectivity 2 and connectivity 3 are almost identical, while connectivity 1 identifies slightly more molecular clouds. This is due to a higher compactness requirement of connectivity 1,  which isolates many small molecular clouds near the edge of large molecular clouds.

\begin{figure}[ht!]
\plotone{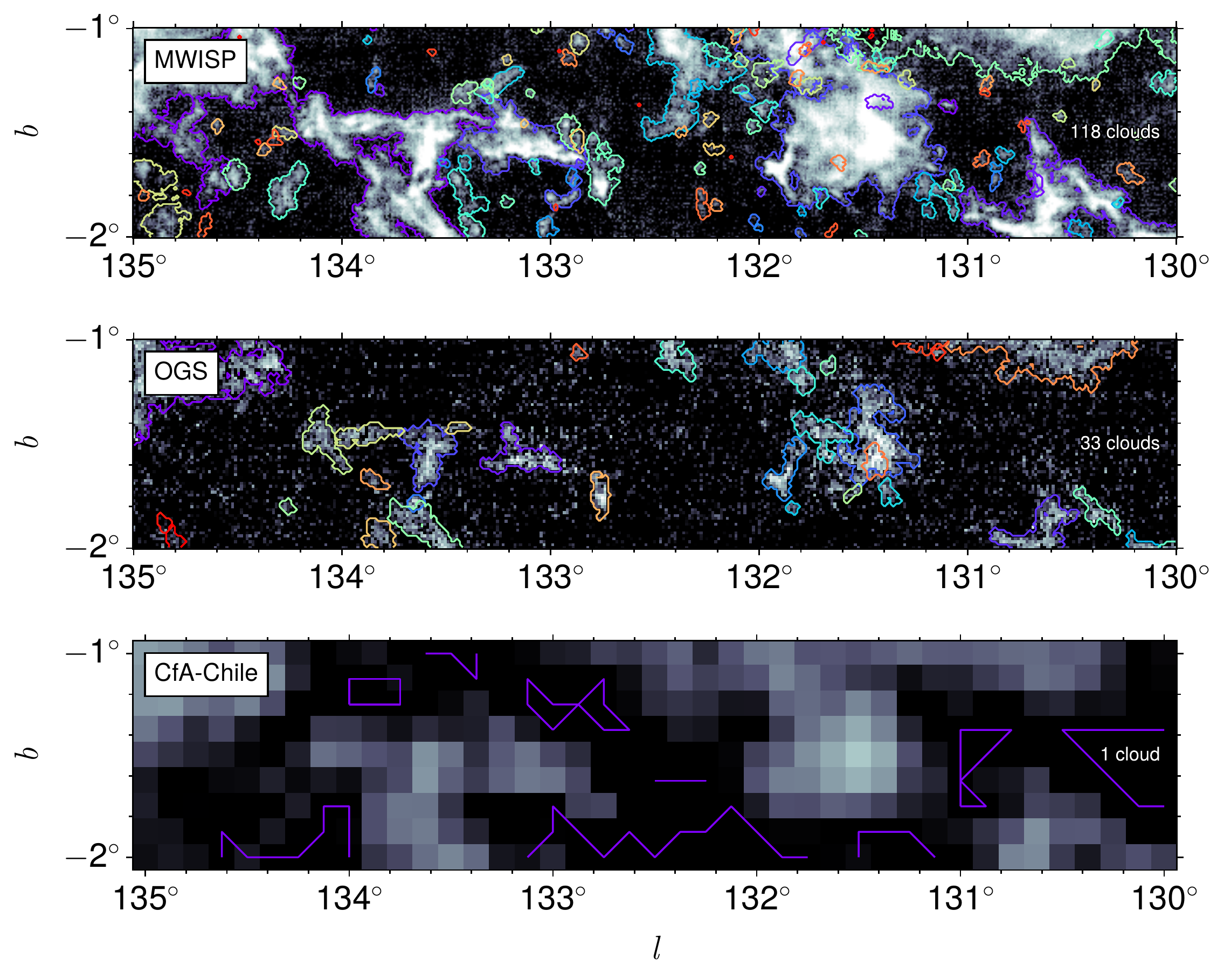}
\caption{Molecular clouds identified in three CO surveys: (top) the MWISP survey, (middle) the OGS, and (bottom) the CfA-Chile survey. In order to reveal details, the demonstrated images zooms in on a part of the PPV data cubes ($130\deg \leq l \leq 135\deg$, $-2\deg \leq b \leq-1\deg$, and $-20 \leq  V_{\rm LSR} \leq  -10$ \kms). The background and color lines are the integrated CO maps and the integrated regions of molecular clouds (defined by DBSCAN) from $-20$ to $-10$ \kms, respectively. Numbers of molecular clouds only take account of those in the above-mentioned PPV range. \label{fig:threesurveyimage}} 
\end{figure}

\subsection{Comparisons between three CO surveys}

\label{sec:threeco}

In this section, we compare molecular cloud samples identified in the three CO surveys with the same DBSCAN parameters. The angular resolution of the MWISP survey is close to that of the OGS, but the sensitivity of the OGS is lower. As to the CfA-Chile survey, both the angular resolution and the sensitivity are significantly lower than the MWISP survey.

 \begin{figure}[ht!]
 \plottwo{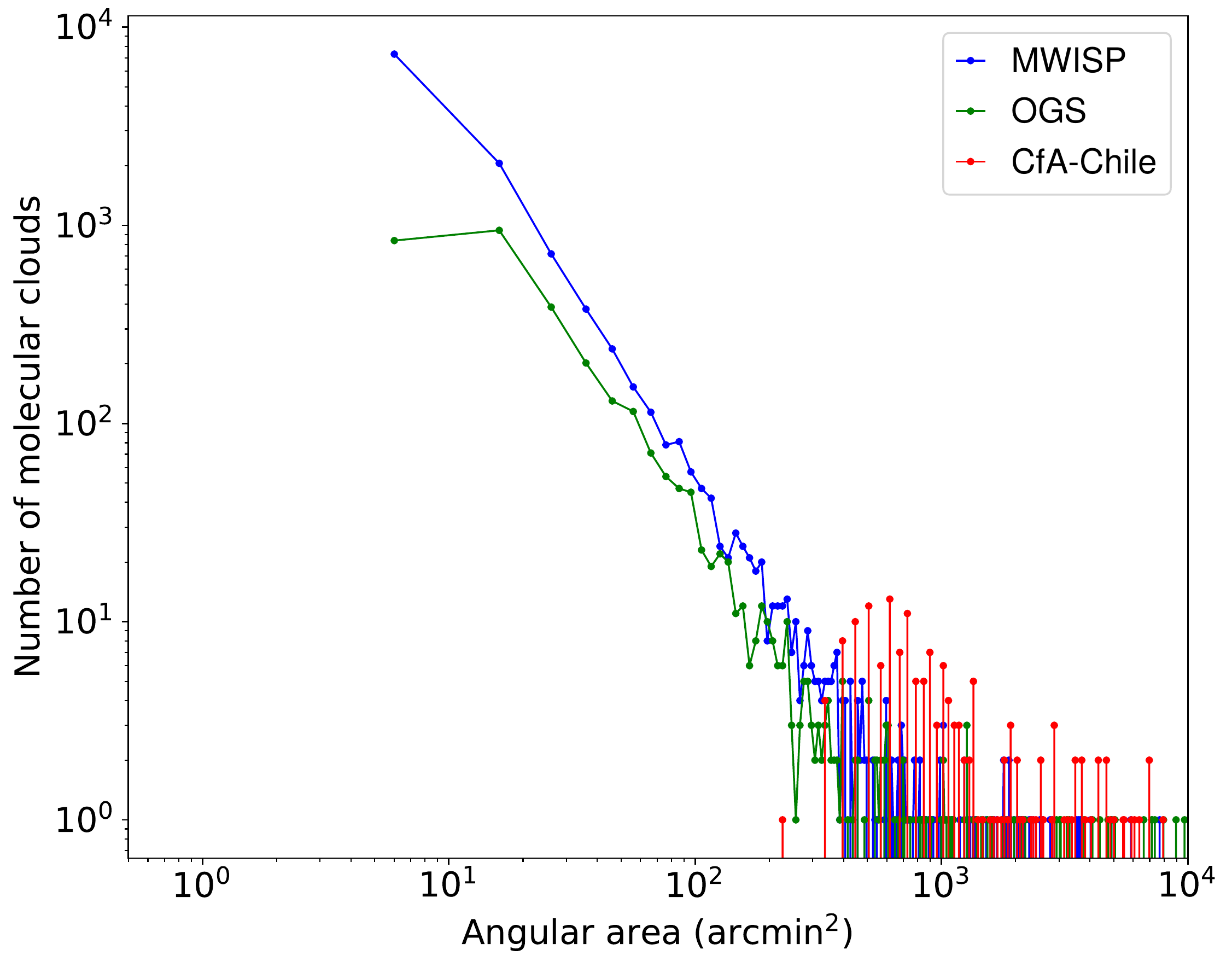}{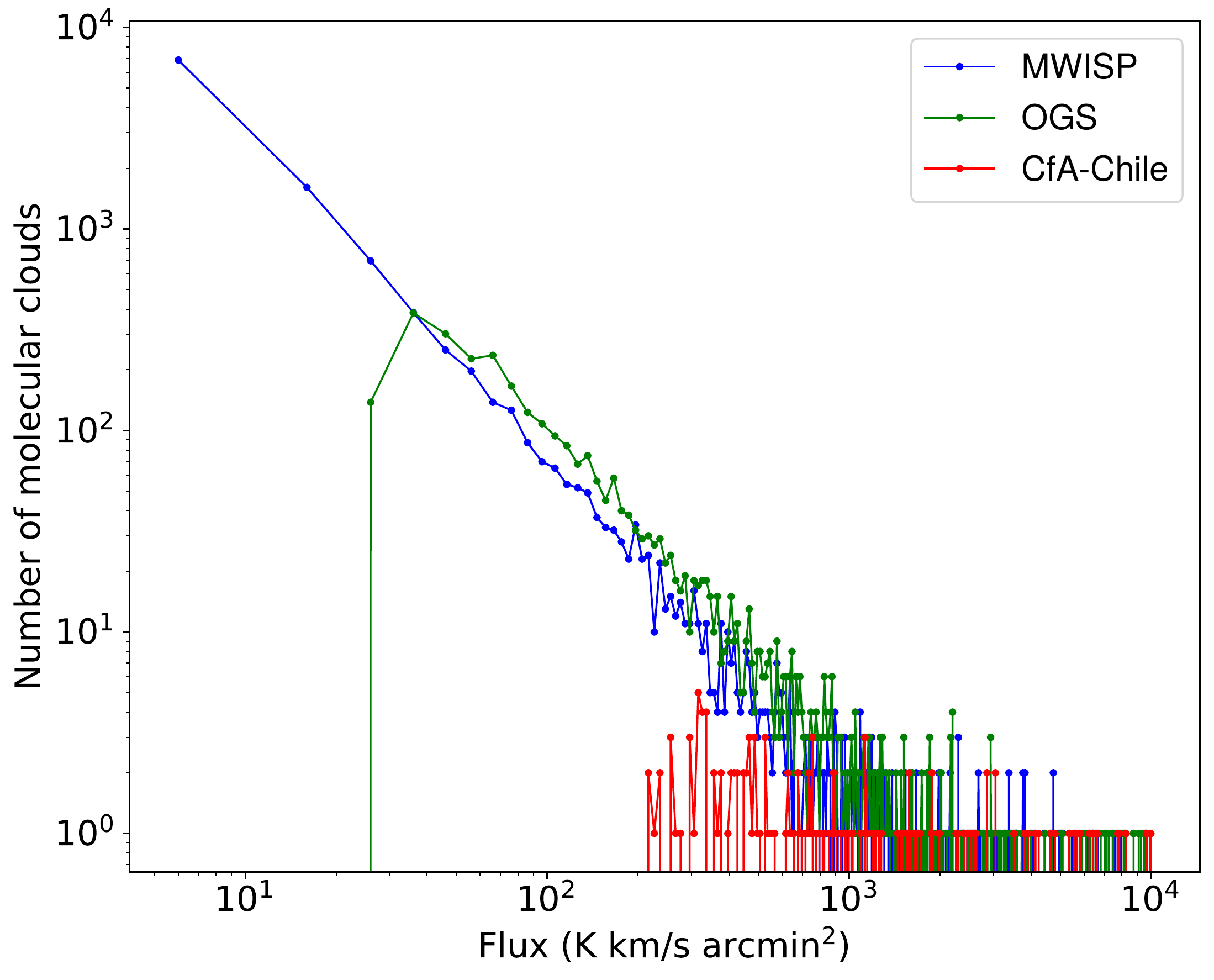}
\caption{Distribution of the angular area and the flux of molecular clouds in three CO surveys: the MWISP survey, the OGS, and the  CfA-Chile survey.\label{fig:s3areaflux}} 
\end{figure}


Figure \ref{fig:threesurveyimage} displays the integrated intensity maps of three CO surveys superimposed by edges of molecular cloud. The intensity is integrated  over $-20 \leq  V_{\rm LSR} \leq  -10$ \kms. Molecular cloud samples produced with three CO surveys are significantly different. In this subregion, the CfA-Chile survey contains only one molecular cloud, while the OGS and the MWISP survey observe 33 and 118 molecular clouds, respectively.



The interplay of angular resolution and sensitivity is evident. Regions that belong to one single molecular cloud in the CfA-Chile survey are split into individuals in the OGS, but some merge into single ones again in the MWISP survey. Generally, large beam sizes merge molecular clouds in PPV data cubes, while low sensitivity splits molecular clouds.

We compare the distribution of the angular area and flux of three surveys in Figure \ref{fig:s3areaflux}. Both the angular area and flux show power-law distributions in the OGS and the MWISP survey, while due to the low angular resolution, distributions of CfA-Chile clouds are heavily deformed. These  power-law distributions resemble those of physical areas and masses \citep{2020ApJ...898...80Y}.

\begin{deluxetable}{ccccccccc}
\tablecaption{GMC parameters. \label{Tab:gmc}}
\tablehead{
\colhead{Name} & \colhead{$l$\tablenotemark{a}} & \colhead{$b$\tablenotemark{a}}  &  \colhead{$V_{\rm LSR}\tablenotemark{a}$}  &  \colhead{Distance} & \colhead{$N_{\rm MWISP/OGS/CfA-Chile}$\tablenotemark{b}}     &  \colhead{$M_{X_{\rm CO}}$\tablenotemark{c}}  & \colhead{$M_{{\rm Virial}}$\tablenotemark{c}}  \\
\colhead{} & \colhead{(\deg)} & \colhead{(\deg)} & (\kms) & (kpc)  &     & ($10^5$\msun) & ($10^5$\msun) &
}

\startdata
 W3  &    (132.43, 134.85)    &   (-0.42, 1.84)  &   (-56.8, -33.7) &  1.95 &  186/60/7  &   2.19/1.89/1.78  &   7.26/4.70/10.27  \\ 
  W5  &   (134.92, 140.27)   &   (-1.27, 2.43)  &  (-44.8, -32.4) & 2.0 &  470/154/11  &      2.08/1.60/3.47  &   7.90/3.27/18.38  \\ 
  NGC 7538  &     (110.27, 112.63)   &  (-0.39, 1.46)  &    (-63.4, -42.6)  & 2.65 &  141/64/1     &  8.30/7.82/9.74  &  19.76/11.50/31.66 \\ 
\enddata 
\tablenotetext{a}{Ranges of $l$ and $b$ correspond to the minimum rectangular boxes that contains the 3$\sigma$ interval along both the major and minor axes determined by \citet{2016ApJ...822...52R}. The $V_{\rm LSR}$ range is the 3$\Delta V$ interval, where $\Delta V$ is the radial velocity dispersion (the second momentum).}

\tablenotetext{b}{Number of molecular clouds in the PPV range of GMCs identified from three CO surveys, the MWISP, the OGS, and the CfA-Chile. } 

\tablenotetext{c}{Masses of molecular clouds with the CO-to-H$_2$ conversion factor, $M_{X_{\rm CO}}$, and the virial equilibrium assumption, $M_{{\rm Virial}}$.  Masses take into account all molecular clouds with centroids in the $l$, $b$, and $V_{\rm LSR}$ ranges. }   

\end{deluxetable}

\section{Individual GMCs}
\label{sec:gmc}

The GMC is one of the most important  molecular cloud populations. Originally, the concept of GMCs was developed from low angular resolution or low spacing observations at early stages of molecular cloud studies \citep[e.g.][]{1979IAUS...84...35S,1987ApJ...319..730S}. However, the definition of GMCs has changed with the progress of observational techniques.

The minimum mass of GMCs are usually considered to be 100,000 \msun\ \citep{1979IAUS...84...35S,1985ApJ...289..373S}, while the diameter is less constrained, could be 10 \citep{1979IAUS...84...35S} or 22 pc \citep{1985ApJ...289..373S}. However, due to the limited angular resolution, GMC catalogs in extra galaxies may only complete down to $10^6$ \msun\ \citep{2021MNRAS.504.6198M}. Given the continuity of molecular cloud mass and size distributions \citep[e.g.][]{2016ApJ...822...52R, 2020ApJ...898...80Y}, however, definitions by mass and size would cause confusion near the thresholds.

Recent studies indicate that there is still no formal definition of GMCs. For instance, \citet{2020ApJ...898....3L} use a threshold of 1000 \msun\ and  6 pc diameter to define GMCs, allowing many more GMC samples in statistical studies. In simulation studies, GMCs are usually defined in a different way. For example, \citet{2013MNRAS.432..653D} used a total column density of 50 \msun\ $\rm pc^{-2}$ to define GMCs, and in terms of star formation, GMCs can also be defined as the largest self-gravitating clouds of the ISM \citep{2020MNRAS.492..488G}.



In this section, we present images of three individual GMCs, described in Table \ref{Tab:threeSurvey}, seen by three CO surveys. The motivation is to examine the variation of GMCs in different observations. The PPV range of GMCs is defined according to their $l$, $b$, and $V_{\rm LSR}$ size (within 3$\sigma$ interval) determined by \citet{2016ApJ...822...52R}.

\begin{figure}[ht!]
 \plotone{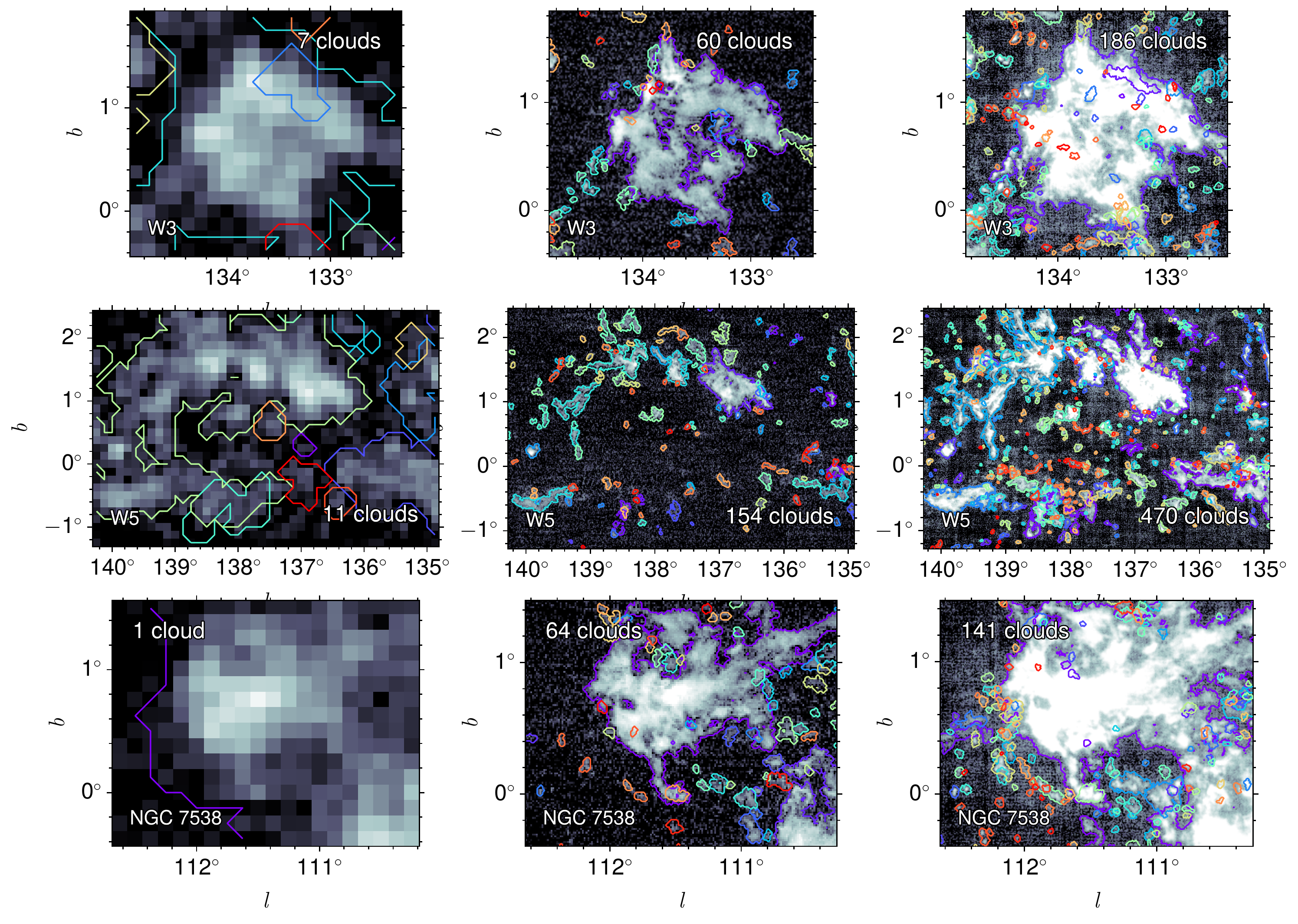}
\caption{Molecular cloud samples of three GMCs, W3, W5, and NGC 7538, in three CO surveys. The left, middle, and right columns show results of the CfA-Chile survey, the OGS, and the MWISP survey, respectively. See Table \ref{Tab:gmc} for the PPV range of GMCs. \label{fig:W3578image}} 
\end{figure}

\begin{figure}[ht!]
 \plotone{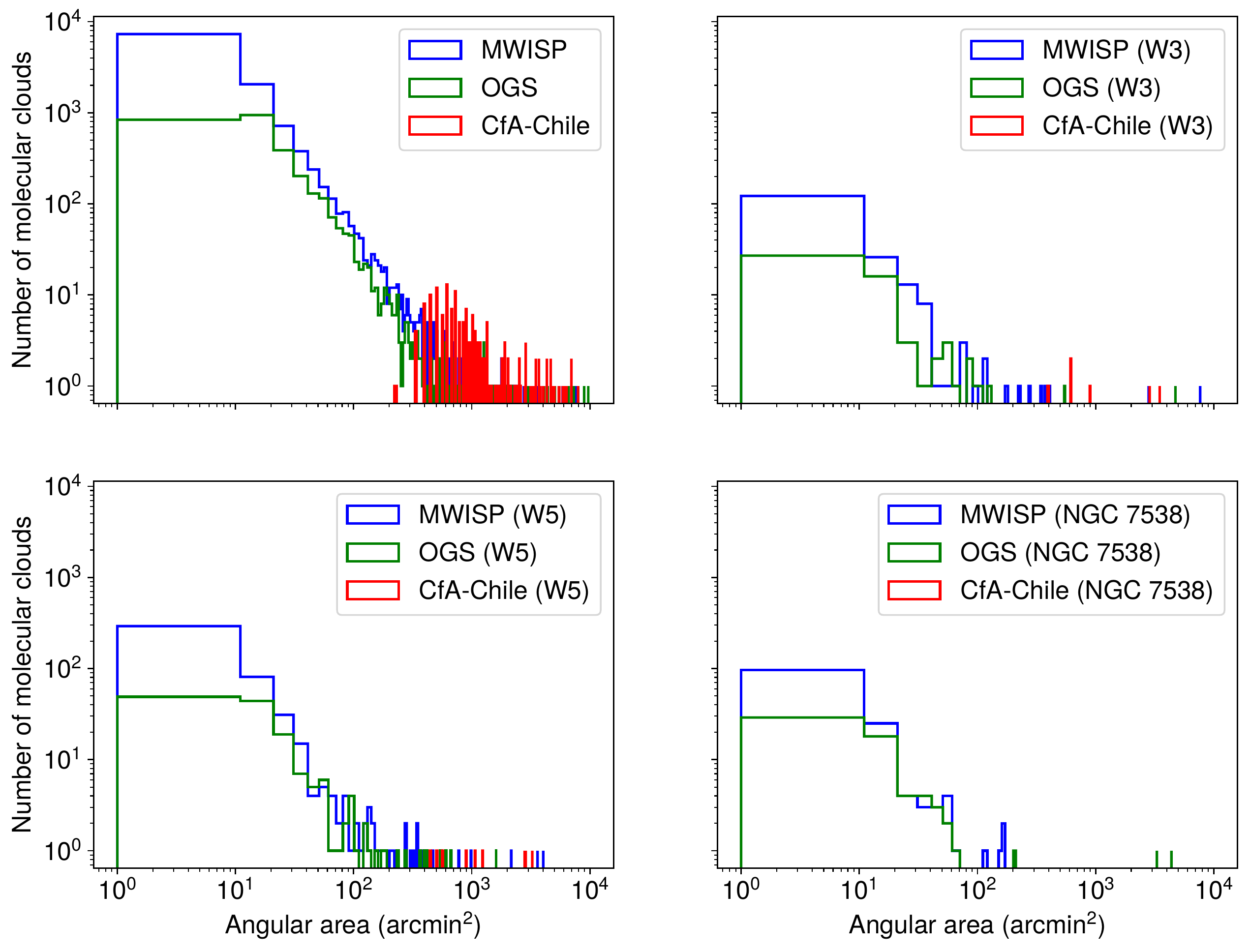}
\caption{Angular area distributions of molecular clouds in three GMCs, W3, W5, and NGC 7538. See Figure \ref{fig:W3578image} for a view of those samples. As a comparison, distributions of the molecular cloud samples in the entire PPV region (see Table \ref{Tab:threeSurvey}) are shown in the upper left panel. \label{fig:W3578area}} 
\end{figure}

\begin{figure}[ht!]
 \plotone{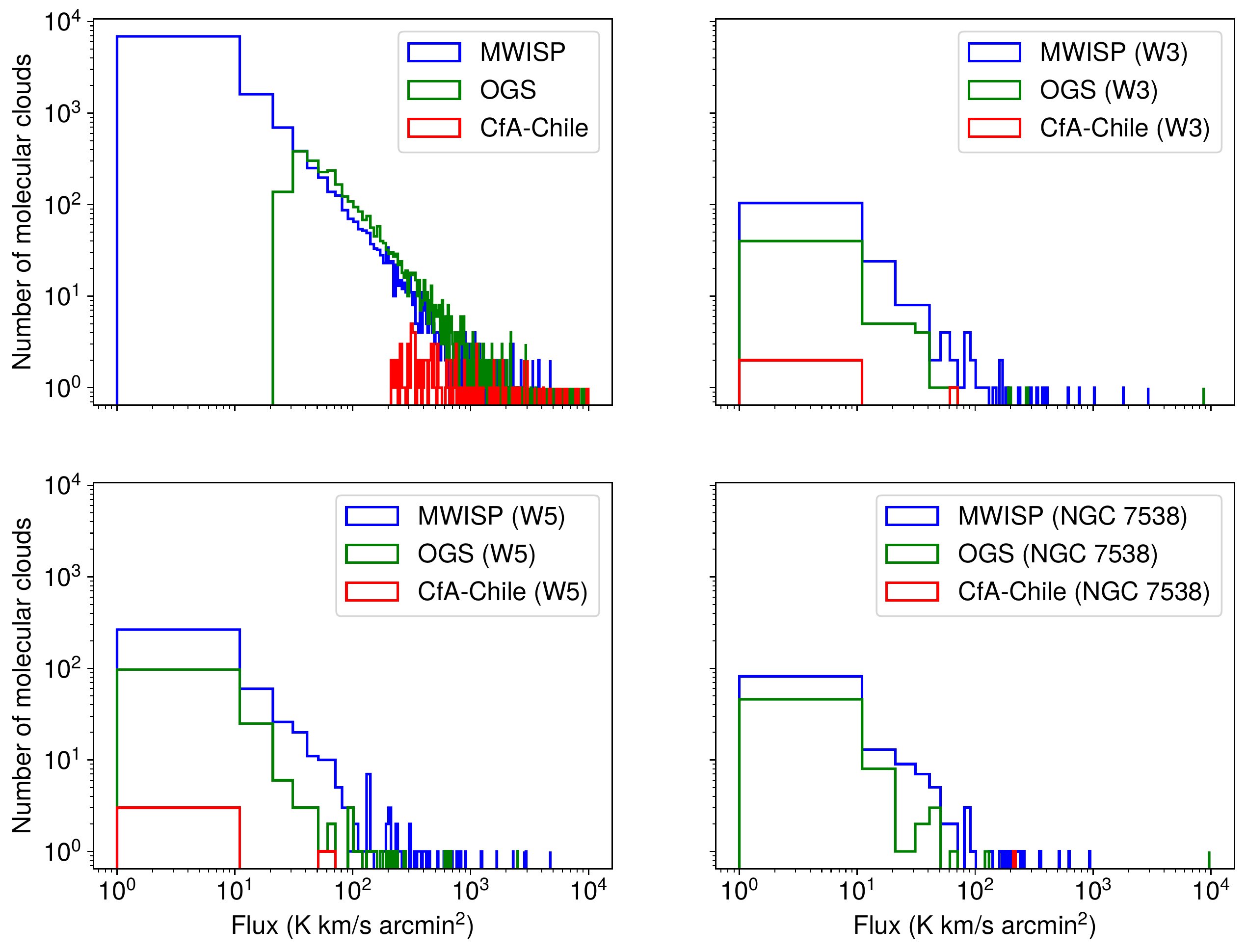}
\caption{Flux distributions of molecular clouds in three GMCs, W3, W5, and NGC 7538. See Figure \ref{fig:W3578image} for a view of those samples. As a comparison, distributions of the molecular cloud samples in the entire PPV region (see Table \ref{Tab:threeSurvey}) are shown in the upper left panel. \label{fig:W3578flux}} 
\end{figure}

The integrated intensity map is displayed in Figure \ref{fig:W3578image}, and distributions of the angular area and flux are  presented in Figures \ref{fig:W3578area} and \ref{fig:W3578flux}, respectively. In Figures \ref{fig:W3578area} and \ref{fig:W3578flux}, as indicated by the MWISP survey and the OGS, distributions of the angular area and flux of GMCs resemble that of molecular clouds in the whole investigated PPV region.

\subsection{W3}

W3 is a high-mass star forming region in the second Galactic quadrant \citep[e.g.][]{2008hsf1.book..264M}, containing four prominent CO components \citep{2011ApJS..196...18B}, W3 Main, W3(OH), AFGL 3333, and W3 North, at a distance of 1.95 kpc \citep{2006Sci...311...54X}. This reservoir of molecular clouds has been mapped using  many molecular lines \citep{1980ApJ...237..711D,1996ApJ...458..561D,1998ApJ...502..265H,2019ApJS..242...19L}, and \cof\ observations reveal a total mass of about $4.4\times10^{5}$ \msun\ \citep{2012MNRAS.422.2992P} for the W3 GMC.

Defining the edge of molecular clouds as contours at 1.4 K based on the OGS, \citet{2001ApJ...551..852H} cataloged W3 as a GMC, giving a diameter of about 128 pc. Using the CfA-Chile survey and an alternative algorithm, however, \citet{2016ApJ...822...52R} derived a diameter of about 83 pc and a mass of $3.5\times10^{5}$ \msun. The assigned index of W3 in the catalog of \citet{2016ApJ...822...52R} is 452, while in the catalog of \citet{2001ApJ...551..852H} is 3646.





To compare with previous results, we calculate the mass of W3 based on the MWISP CO survey.  Converting the \cofs\ flux to the mass by the CO-to-H$_2$ conversion factor (X-factor), $X_{\rm CO} = 2\times10^{20} \ \rm cm^{-2} \ (K \ km s^{-1})^{-1}$ \citep{2013ARA&A..51..207B}, the mass of W3 is about $2.19\times10^{5}$ \msun\ (see Table \ref{Tab:gmc}), which is significantly smaller than its virial mass, $7.26\times10^{5}$ \msun.






\subsection{W5}

W5 is another GMC that is not far from W3, and it consists of two main components, W5 West and W5 East \citep{2003ApJ...595..900K}. Its distance is about 2 kpc \citep{2012A&A...546A..74D} and it is located near the open cluster IC 1848 \citep{2014MNRAS.438.1451L}. This region has been used to study the feedback of high-mass stars on their natal molecular clouds \citep{2003ApJ...595..900K,2011MNRAS.418.2121G,2019ApJS..242...19L}.

Based on previous studies, the total mass of molecular gas in W5 was about $5\times10^{4}$ \msun\ \citep{2011MNRAS.418.2121G,2003ApJ...595..900K}, and was about  $6\times10^{4}$ \msun\ if the atomic gas is also included \citep{2008ApJ...688.1142K}. In the catalog of \citet{2016ApJ...822...52R}, the index of W5 is 515, and its mass about $1.38\times10^{5}$ \msun\ based on a kinematic distance of 2.75 kpc.

However, in the MWISP survey, the $X_{\rm CO}$ mass of the W5 is about $2.08\times10^{5}$ \msun\ and the virial mass ia about $7.9\times10^{5}$ \msun, satisfying the criterion of GMCs. Among those three GMCs, W5 is the most fragmented one seen by the MWISP survey, resulting 470 molecular clouds.

\subsection{NGC 7538}

NGC 7538 is the natal molecular cloud of a well-known high-mass protostar NGC7538IRS1. The mass of this embedded protostar  is about 30 \msun\ \citep{1984ApJ...282L..27C}, containing  both 12 GHz and 6 GHz methanol masers \citep{2000A&A...362.1093M}, at a distances of about 2.65 kpc \citep{2009ApJ...693..406M}.

\citet{1981ApJ...250L..43D} derived a minimum mass of $8\times10^{4}$ \msun\ for the molecular mass in the GMC NGC 7538. As they have predicted, the actual mass is much higher, and is about $7.94\times10^{5}$ \msun\ \citep[index 418 in the catalog of][]{2016ApJ...822...52R}.

In this work, the MWISP survey collects an $X_{\rm CO}$  mass of about  $8.30\times10^{5}$ \msun\ and suggests a virial mass of $2.0\times10^{6}$ \msun. Interestingly, NGC 7538 is more massive than both W3 and W5, but it contains the smallest number of molecular clouds.

\section{Discussion}
\label{sec:discuss}

\subsection{Sensitivity}

Sensitivity is essential in the building of molecular cloud samples. First of all, sensitivity determines the level of cutoffs in PPV data cubes, and the mean voxel SNR of molecular clouds is strongly related to the observed fraction of the total flux \citep{2021ApJ...910..109Y}.

In terms of the molecular cloud number, sensitivity has two effects. First, high sensitivity expands the outskirts of molecular clouds and connects those molecular clouds seen isolated at low sensitivity. Secondly, high sensitivity also reveals small faint molecular clouds that are not seen at low sensitivity. Consequently, the detected molecular cloud number is an interplay between sensitivity and the size distribution of molecular clouds. Based on the power-law size distribution and the results of the MWISP survey and the OGS, the amount of small faint molecular clouds is significant.

In Figures \ref{fig:sensitivityPlot}, \ref{fig:smMWISPplot}, and \ref{fig:velsmPlot}, numbers of molecular clouds in the Outer and OSC arms are significantly lower than that of the Local and Perseus arms. Consequently, due to low beam filling factors, high sensitivity is needed for observations of distant molecular clouds. Clearly,  molecular cloud samples in the Outer and OSC arms are incomplete.

High sensitivity also increases the angular area of molecular clouds.  In addition to direct comparisons in Figures \ref{fig:sensitivityPlot} and \ref{fig:sensitivityAreaSingle},  distribution as of angular area (Figure \ref{fig:s3areaflux}) also show that the angular area of molecular clouds in the MWISP CO survey is systematically larger than that in the OGS. This angular area dependence on resolution and sensitivity calls for caution when molecular cloud samples from multiple surveys are used collectively. Properties of molecular clouds in observations obtained with different telescopes are usually affected by sensitivity, resolution, or even the calibration routine \citep{2015ApJ...814...83L,2019MNRAS.483.4291C}. Consequently, statistics with hybrid samples may contain systematic errors caused by observations. 


\subsection{Effects of Angular and Velocity Resolution}

Angular resolution affects molecular cloud samples in two ways as mentioned in section \ref{sec:ang}, and both yield significant effects.  First, low angular resolution imposes large dilution factors (beam filling factors) on the brightness temperature \citep{2021ApJ...910..109Y}, and distant objects are observed to be faint. Consequently, for those molecular clouds, high sensitivity are needed to collect equivalent amount of flux, and many small molecular clouds are also missed due to low peak intensities in PPV space. Secondly, low angular resolution merge molecular clouds in PPV space and decrease the voxel number of molecular clouds in PPV data. As demonstrated in Figure \ref{fig:threesurveyimage}, this effect is obvious, and the CfA-Chile survey identifies only one molecular cloud in the shown subregion. In this case, many small molecular clouds are excluded due to the requirement of a minimum voxel number of 16.

The CfA-Chile survey of the Milky Way resembles high angular resolution observations toward molecular clouds in galaxies of the Local Group, both characterized with low beam filling factors. For instance, \citet{2016A&A...585A..44M} found an averaged beam filling factor of 0.8\% for dense gas in the Andromeda (M31) with IRAM 30 m radio telescope. Considering a distance of 780 kpc of M31, a 1.2 m telescope used by the CfA-Chile survey would yield a similar physical resolution for molecular clouds at 31 kpc. For dense gas at about 5 kpc, the averaged beam filling factor for the CfA-Chile survey is estimated to be about 0.3.

 Low beam filling factors could cause significant effects on extragalactic studies. For instance, \citet{2020MNRAS.495.3819A} found that a certainly level of resolution (3-4 telescope beams per galaxy scale) and signal-to-noise ratio (SNR) is required to obtain accurate metallicity gradient measurements. The usage of $\mathrm{CO}\left(J=2\rightarrow1\right)/\mathrm{CO}\left(J=1\rightarrow0\right)$ line ratio \citep[e.g.,][]{2014A&A...567A.118D,2021MNRAS.504.3221D,2022ApJ...926...96M} is an additional example of angular resolution effects. Given the uncertainty of beam filling factors, the derived GMC mass with $\mathrm{CO}\left(J=2\rightarrow1\right)$ line may contain large systematic errors. This beam filling factor effect is also applicable to various extragalactic studies, such as gas surface density \citep[e.g.,][]{1989ApJ...344..685K,2002ApJ...569..157W, 2008AJ....136.2846B}, dense gas fraction \citep[e.g.][]{2004ApJ...606..271G, 2007ApJ...660L..93G}, star formation rate \citep[see][for a review]{2012ARA&A..50..531K}, and star formation efficiency \citep[e.g.,][]{2004ApJ...606..271G, 2018ApJ...860..165T}.  



The angular resolution effect indicates that samples that contain molecular clouds at different distances are not uniform. For instance, \citet{2010ApJ...723..492R} discussed the Malmquist bias in molecular clouds identified in the GRS. The Malmquist bias is caused by the decreasing of completeness toward far distances. This non-uniformity may cause systematic statistical errors. 


Velocity resolution, however, only affects the PPV data in one axis, so the effect is similar but less severe compared with angular resolution. Due to the requirement of the minimum velocity channel number of 3, small molecular clouds are missed if velocity resolution is insufficient.

\begin{deluxetable}{ccccccccccc}
\tablecaption{Algorithm parameters. \label{Tab:para}}
\tablehead{
\colhead{Algorithm} & \colhead{MinPts} & \colhead{Connectivity}  &  \colhead{min\_delta}  &  \colhead{Decomposition by} & \colhead{ Cutoff}     &  \colhead{Voxel\tablenotemark{a}}  & \colhead{Has a beam} & \colhead{Channel\tablenotemark{a}} & \colhead{Peak\tablenotemark{a}}  \\
\colhead{} & \colhead{(\deg)} & \colhead{(\deg)} & (\kms) & (kpc)  &     & ($10^5$\msun) & ($10^5$\msun) &
}

\startdata
 DBSCAN  &   4    &   1  &  --  &   --  &                3$\sigma$  &  16  & Yes & 3 & 5$\sigma$\\ 
  HDBSCAN  & variable   &  variable &  --  &  --   &     3$\sigma$  &  16  & Yes & 3 & 5$\sigma$\\ 
  SCIMES  &    --   &  1  &    3$\sigma$  &  volume &    3$\sigma$  &  16  & Yes & 3 & 5$\sigma$\\ 
\enddata 
\tablenotetext{a}{The minimum number or value required for a molecular cloud.}  

\end{deluxetable}

\subsection{DBSCAN parameters}

 In PPV space, molecular clouds are in essence defined by algorithms. \citet{2015MNRAS.454.2067C} compared molecular cloud samples identified with four algorithms, SCIMES, Dendrogram, \texttt{CPROPS}, and \texttt{CLUMPFIND}. Both the number and edge of molecular clouds vary significantly, and the results would be certainly different if the DBSCAN algorithm was used. 

 The essential difference between DBSCAN and other algorithms is that DBSCAN identifies isolated objects and ignores their internal structures in PPV space, while other algorithms are likely to split DBSCAN clouds into multiple objects.  The splitting criteria is somewhat subjective, and in this work, we only examine the variation of molecular clouds with respect to DBSCAN parameter combinations.


Molecular cloud samples are stable against DBSCAN parameters. However, it is possible that some DBSCAN structures are composed of objects at different distances. Confirming DBSCAN structures needs accurate distance measurements, and we cannot rule out that large structures in PPV space are indeed single molecular clouds. Consequently, distance measurements are essential to verify molecular cloud samples identified in PPV space.  
 
\subsection{Algorithm dependence}

The algorithm used in this work to determine the molecular cloud boundary is DBSCAN. However, DBSCAN is not the unique choice, and in order to see the variation of molecular cloud samples with respect to algorithms, we compare DBSCAN with another two methods, HDBSCAN and SCIMES.

HDBSCAN and SCIMES are both hierarchical clustering algorithms, taking internal structures of independent PPV components into consideration. SCIMES is particularly designed to decompose dendrogram trunks, but decomposing PPV components is the favorite operation of large scale surveys \citep{2015ARA&A..53..583H}. In addition, the criteria of decomposing PPV components is complicated, usually involving many parameters. Although HDBSCAN is an improved version of DBSCAN, due to the variation of connectivity, the definition of molecular clouds in HDBSCAN is non-uniform \citep{2020ApJ...898...80Y}.

We select a local segment ($130\deg <l< 141\fdg54$, $-2\deg < b < 4\deg$, and $-20 < V_{\rm LSR} < 10\ \rm km\ s^{-1}$) to compare three algorithms. Processing the whole PPV region shown in Table \ref{Tab:threeSurvey} with HDBSCAN and SCIMES is slow, and even for this local region, HDBSCAN and SCIMES are still computationally expensive. So we use a cutoff of 3$\sigma$, rather than a usual value of 2$\sigma$.

In HDBSCAN, the minimum cluster size is 16, consistent with DBSCAN settings, and we leave other parameters unchanged. In SCIMES, we segment dendrogram trunks with the \texttt{volume} criteria and save all isolated and unclustered dendrogram leaves. For the convenience of comparison, we recalculate properties of molecular clouds according to the PPV labels provided by HDBSCAN and SCIMES with the same procedure of DBSCAN, and four post criteria (see section\ref{sec:cloudIdentification}) also applied. See Table \ref{Tab:para} for a summary of parameters used in these three algorithms.

\begin{figure}[ht!]
\plotone{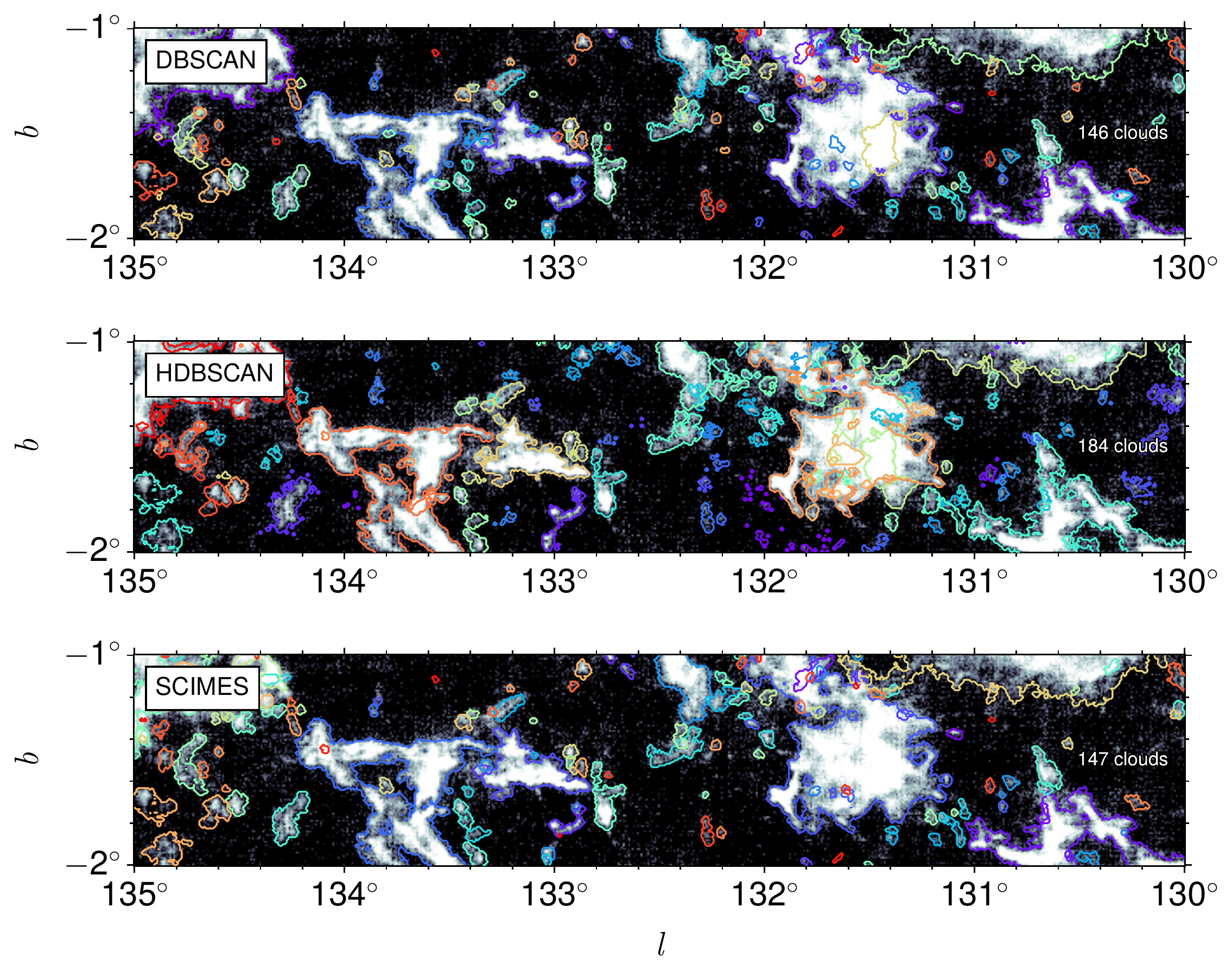}
\caption{Molecular cloud edges defined with three algorithms: DBSCAN (top), HDBSCAN (middle), and SCIMES (bottom). In order to display edges clearly, this demonstration zooms in on a small PPV range ($130\deg \leq l \leq 135\deg$, $-2\deg \leq b \leq-1\deg$, and $-20 \leq  V_{\rm LSR} \leq  -10$ \kms). The background images and color lines are the integrated CO maps and projection edges, respectively. Numbers of molecular clouds only take account of those in the above-mentioned PPV range. \label{fig:threealgorithmimage}} 
\end{figure}

\begin{figure}[ht!]
 \plotone{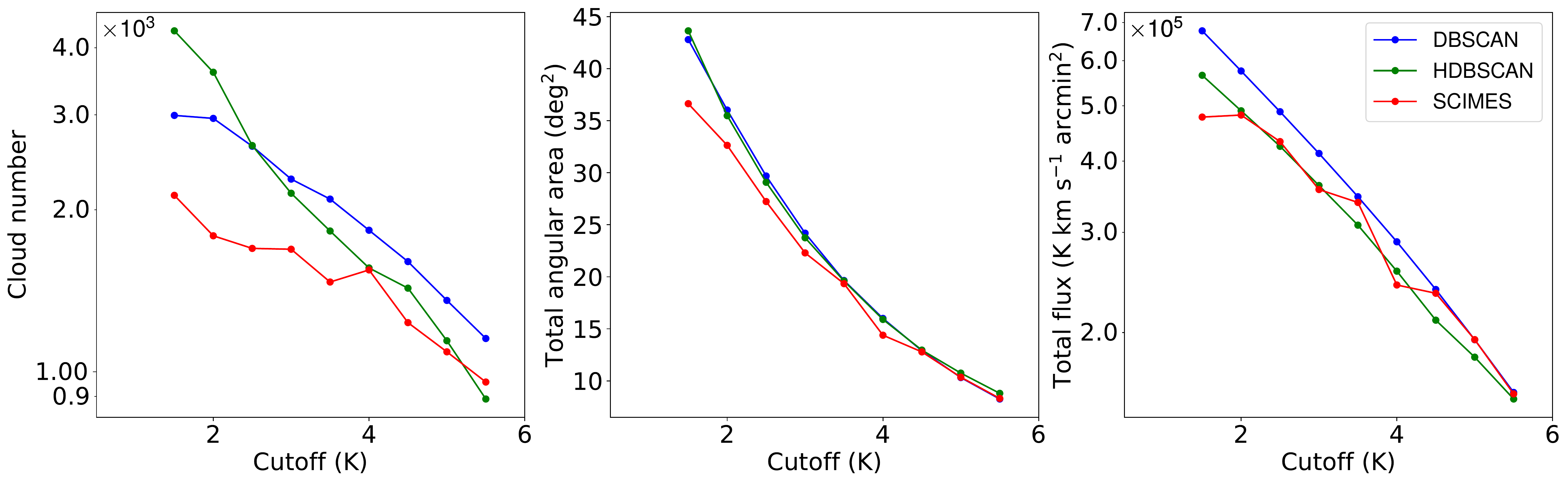}
\caption{Variation of molecular cloud properties with respect to algorithms and sensitivity (cutoffs). From left to right, this demonstration shows the number, the total angular area, and the total flux of molecular clouds identified with three algorithms, DBSCAN (blue), HDBSCAN (green), and SCIMES (red). \label{fig:sensAreaAlgorithm}} 
\end{figure}

\begin{figure}[ht!]
 \plotone{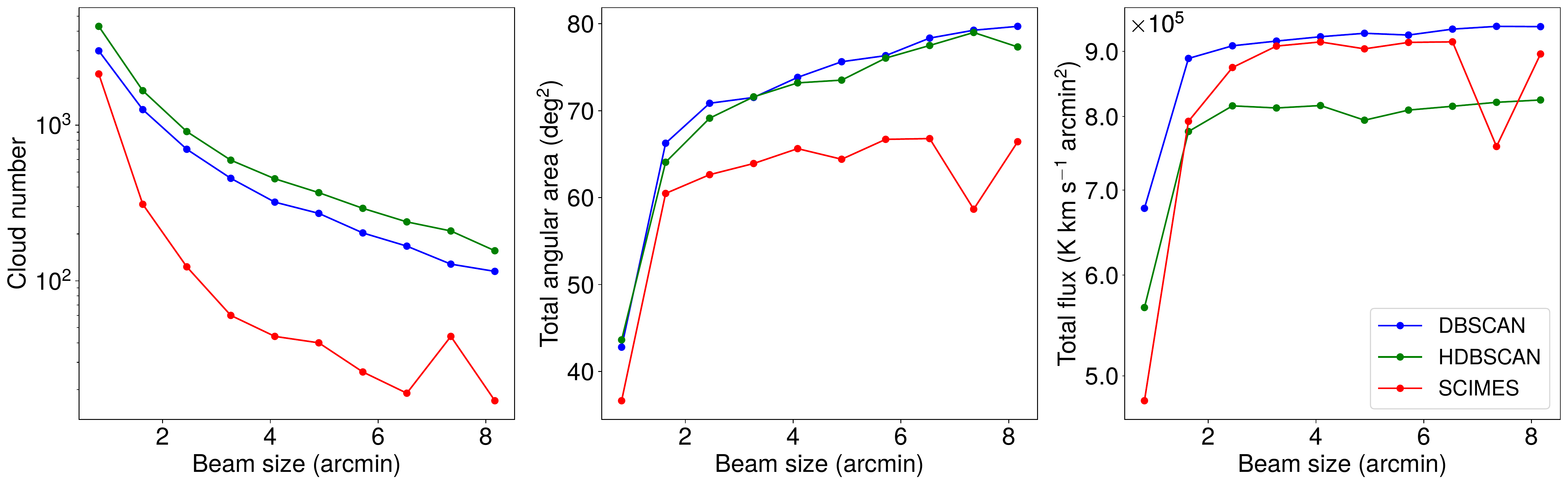}
\caption{Variation of molecular cloud properties with respect to algorithms and  beam sizes (angular resolution). From left to right, this demonstration shows the number, the total angular area, and the total flux of molecular clouds identified with three algorithms, DBSCAN (blue), HDBSCAN (green), and SCIMES (red).   \label{fig:beamAreaAlgorithm}} 
\end{figure}

\begin{figure}[ht!]
 \plotone{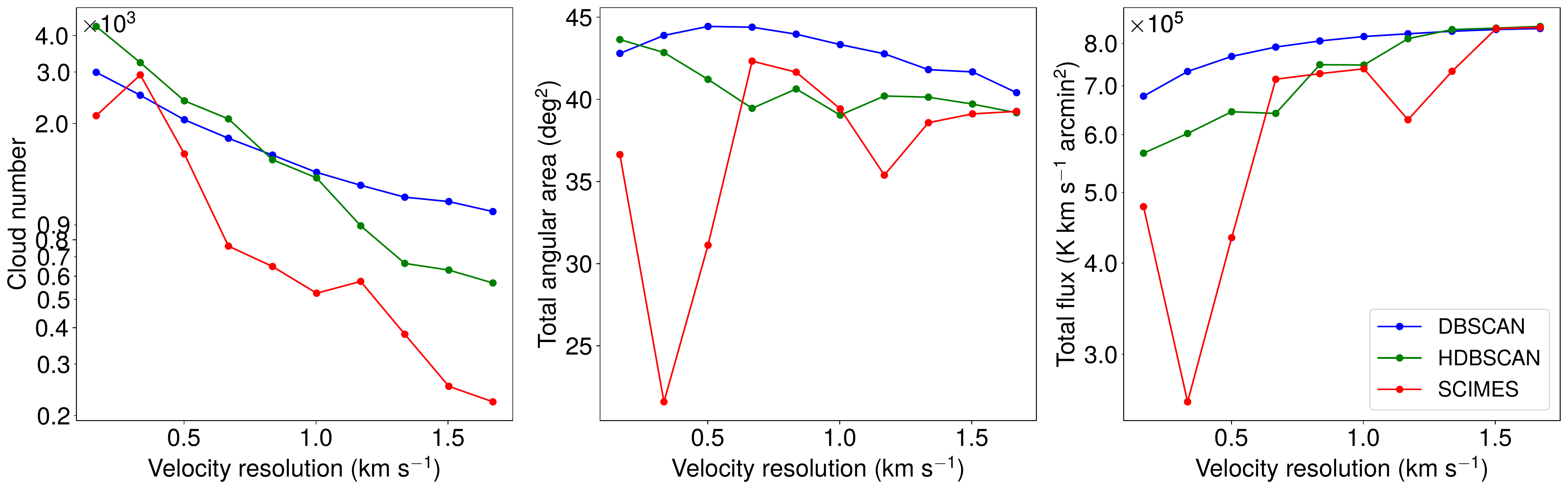}
\caption{Variation of molecular cloud properties with respect to algorithms and velocity resolution. From left to right, this demonstration shows the number, the total angular area, and the total flux of molecular clouds identified with three algorithms, DBSCAN (blue), HDBSCAN (green), and SCIMES (red).  \label{fig:veloAreaAlgorithm}} 
\end{figure}

Figure \ref{fig:threealgorithmimage} zooms in on a small region to display edges of molecular clouds. Figure \ref{fig:sensAreaAlgorithm}-\ref{fig:veloAreaAlgorithm} demonstrate variations of the cloud number, the total angular area, and the total flux with respect to sensitivity, angular resolution, and velocity resolution, respectively.

The difference bettween DBSCAN and HDBSCAN is insignificant, and SCIMES also shows similar trending but with slightly larger fluctuations. HDBSCAN occasionally takes a collection of isolated points as molecular clouds, which is caused by the variation of linkages. Fluctuations in SCIMES, however, is mostly caused by the interplay between sensitivity/resolution and decomposing operations.


Consequently, we use DBSCAN to define molecular clouds in PPV space. First, DBSCAN has only two parameters, and it is the simplest algorithm with uniform definition across all PPV space. Secondly, DBSCAN is robust against the noise, sensitivity, and angular/velocity resolution. Thirdly, through adopting shifting and summing operations, the speed of DBSCAN is fast for large data cubes.

\subsection{Velocity Crowding}

 Velocity crowding aka radial velocity confusion is an intrinsic problem of PPV data, and contiguous structures in PPV space may not represent actual PPP molecular clouds. For instance, \citet{2000ApJ...532..353P} studied the project effects on the morphology of molecular clouds, and they concluded that compared with the density filed, channel maps are more representative of the radial velocity distribution. \citet{2010ApJ...712.1049S}  presented a thorough investigation of the PPV projection effects on the mass-size and linewidth-size relationships. Discrepancies of those relationships in PPV and PPP spaces are significant, and in a specific velocity range, 2D integrated maps contain contributions from gases that have close radial velocity. \citet{2019MNRAS.490.2648B} found that even a low fraction of gas contamination along lines of sight can increase the index of the mass-size relation by 0.2-0.3.   

 \citet{2018MNRAS.479.1722C} also emphasized that coherent structures in PPV space may be physically unrelated. Along lines of sight, there can be multiple components \citep[e.g.,][]{2020ApJ...892L..32S}, and large velocity dispersion due to bulk motions may connect independent molecular clouds in PPV space, degrading the quality of molecular cloud samples. Consequently, mapping from PPV space to PPP space is not straightforward.  For instance,  \citet{2021ApJ...919L...5B} found a spurious filamentary connection between Taurus and Perseus. 

 Importantly, \citet{2022ApJ...925..201P} showed that velocity crowding effects can cause illusion of spiral arms on the Galactic longitude-velocity (LV) diagram. They find that molecular gas in the Perseus Arm is not continuous. This indicates that distance measurements are essential in the study of Galactic structures, and LV diagrams may provide erroneous results.  

\begin{figure}[ht!]
 \plotone{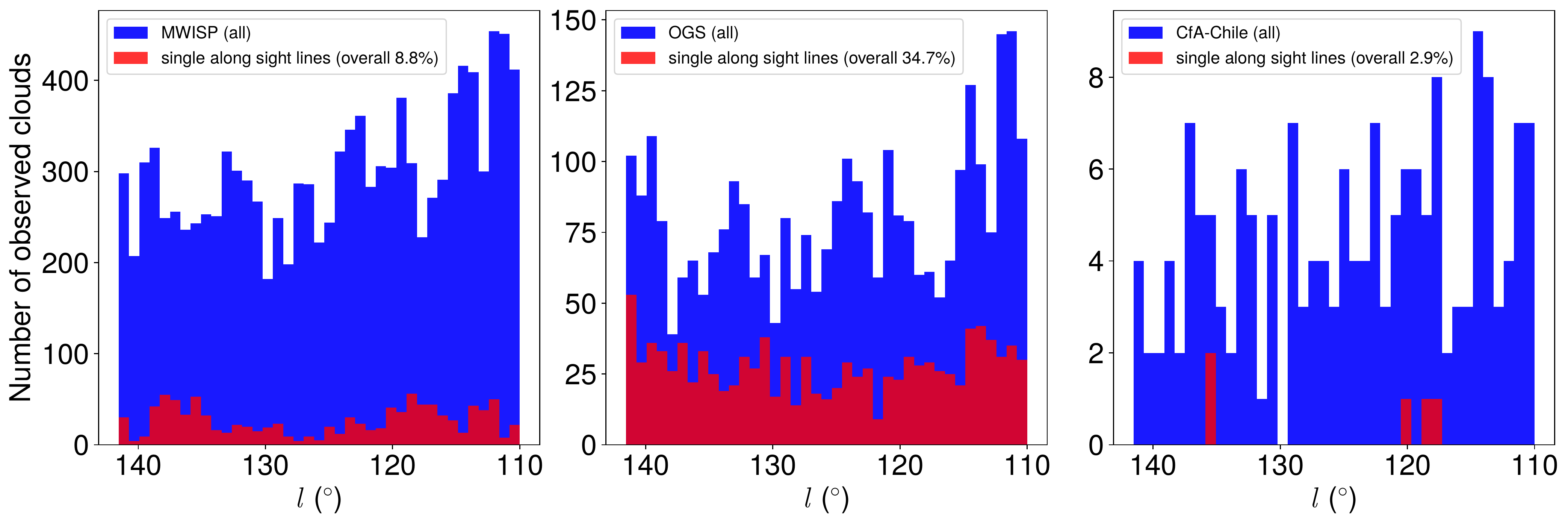}
\caption{Distribution of molecular clouds along the Galactic longitude and portion of clouds that are single along lines of sight. Overall, the portion is 8.8\%, 34.7\%, and 2.9\% for the MWISP survey, OGS, and CfA-Chile survey, respectively. \label{fig:lhist}} 
\end{figure}

 In order to measure the velocity confusion effect on molecular cloud samples, we derive the portion of molecular clouds that have no overlapping ones  along sightlines (toward all voxels). In this examined region, the portion of MWISP is about 8.8\% (see Figure \ref{fig:lhist}), indicating that the velocity projection effect is still significant in the second Galactic quadrant. Regions with high portions  are valuable for molecular cloud studies.   

 \begin{figure}[ht!]
 \plotone{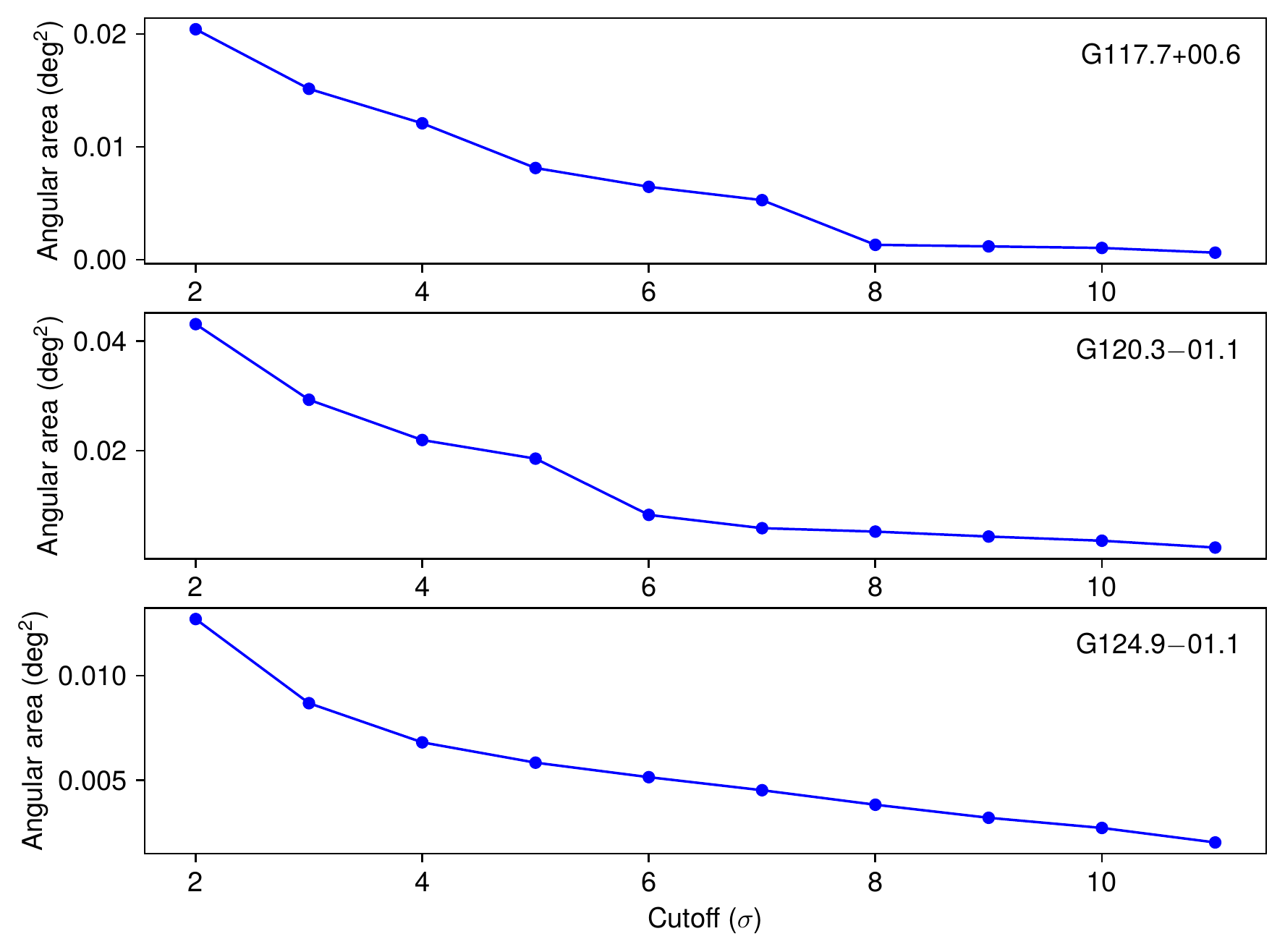}
\caption{Variation of angular areas with respect to cutoffs. These three molecular clouds are unique along their lines of sight. The cutoff is used as a proxy of sensitivity, and high cutoffs mean low sensitivity. \label{fig:mccutoff}} 
\end{figure}

\subsection{Boundary of Molecular Clouds}


 Using a linear extrapolation, \citet{1980ApJ...241..676B} estimated the variation of angular area against sensitivity. They came to a conclusion that the area of molecular clouds would converge at a specific high sensitivity, indicating fairly sharp boundaries. However, as shown in Figures  \ref{fig:sensitivityPlot} and \ref{fig:sensitivityAreaSingle}, a linear extrapolation may underestimate the molecular cloud angular area.

 \citet{1999ApJ...515..286B} proposed an alternative picture of molecular clouds. In this scenario, molecular clouds are essentially density fluctuations due to large-scale interstellar turbulence and are formed during the colliding of gas streams. Consequently, defining cloud boundaries as density-threshold has little physical meaning, i.e., the boundary of molecular clouds is soft or smooth.

 The edge feature revealed by the MWISP survey is consistent with the picture of turbulent density fluctuations \citep{1999ApJ...515..286B}.  Both the total (Figure \ref{fig:sensitivityPlot}) and individual (Figure \ref{fig:mccutoff})  angular area of molecular clouds shows no sign of convergence toward high sensitivity (low cutoff $\sigma$), denying the existence of sharp boundaries. This scenario is in line with recent simulations \citep{2022MNRAS.512.4765S} and observations \citep{2019A&A...623A..16S}.




 \begin{figure}[ht!]
 \plottwo{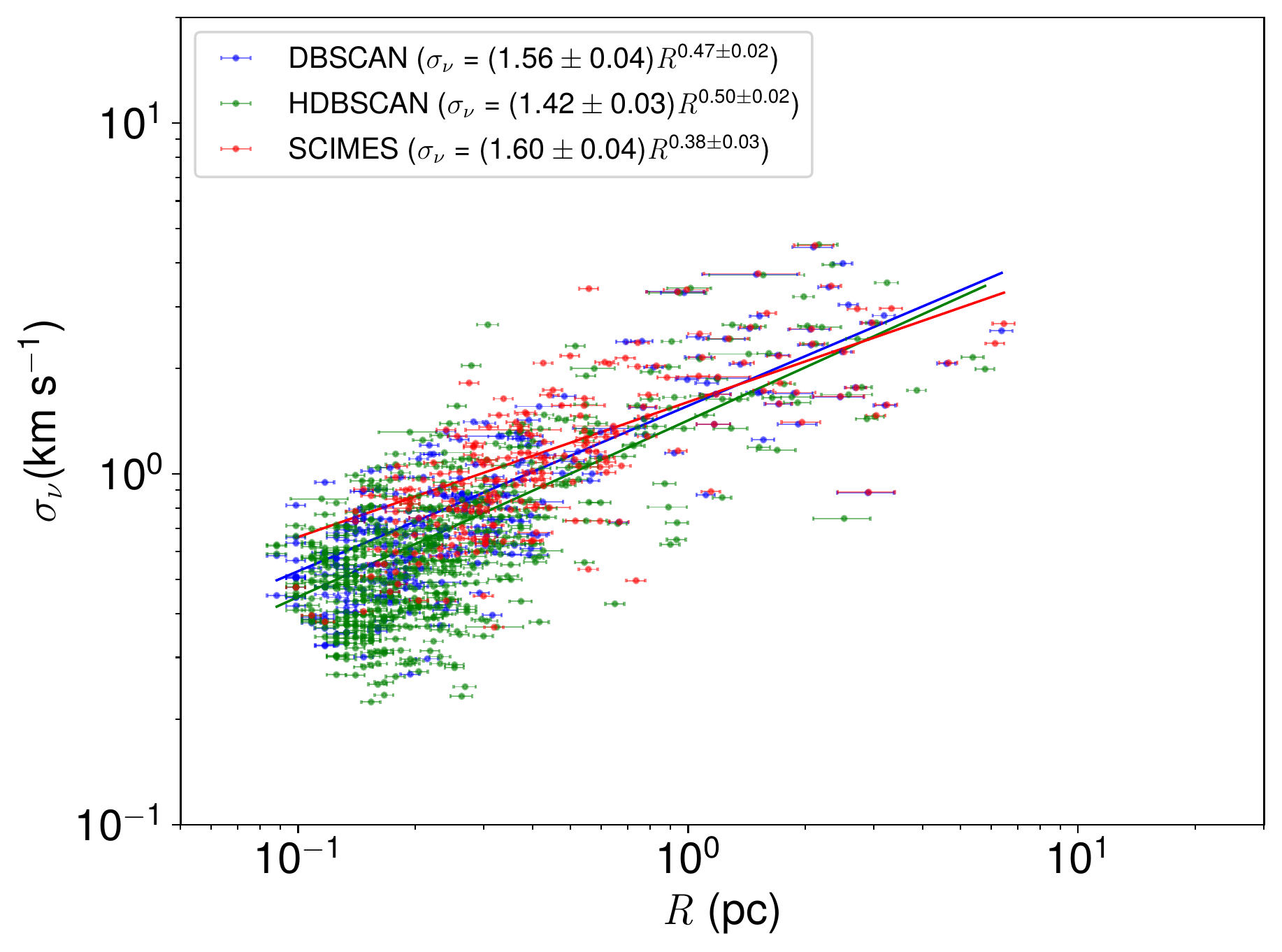}{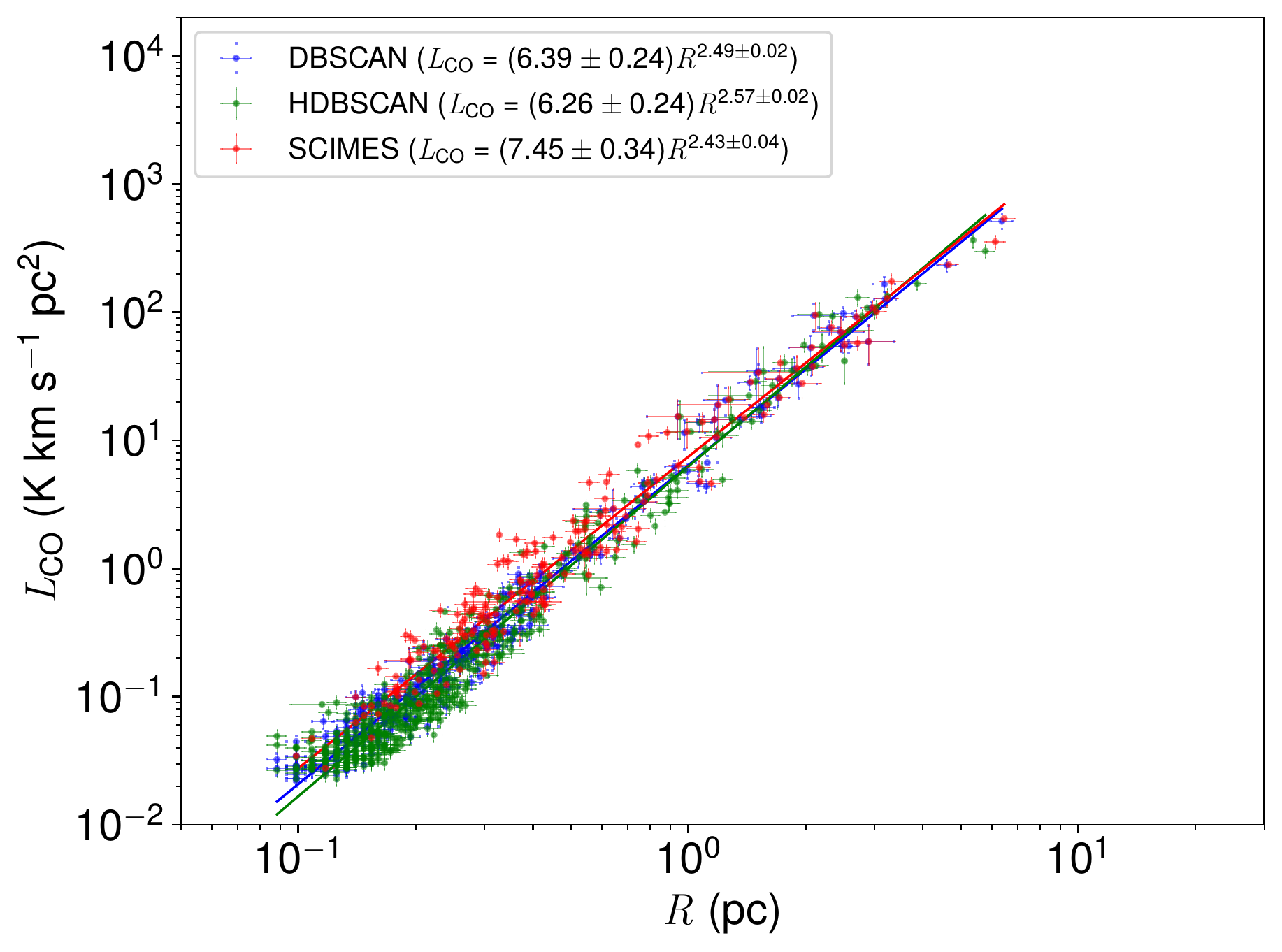}
\caption{Scaling relations between the equivalent line width ($\sigma_{v}$) and radius ($R$, left) and  between the CO luminosity ($L_{\rm CO}$) and radius (right). Results of different algorithms,  DBSCAN (blue), HDBSCAN (green), and SCIMES (red), are color-coded. The cutoff of data cubes is 3$\sigma$, and only errors of distances are considered, including the 5\% systematic error \citep{2021ApJ...922....8Y}. \label{fig:larsonAlgo}} 
\end{figure}

 \begin{figure}[ht!]
 \plottwo{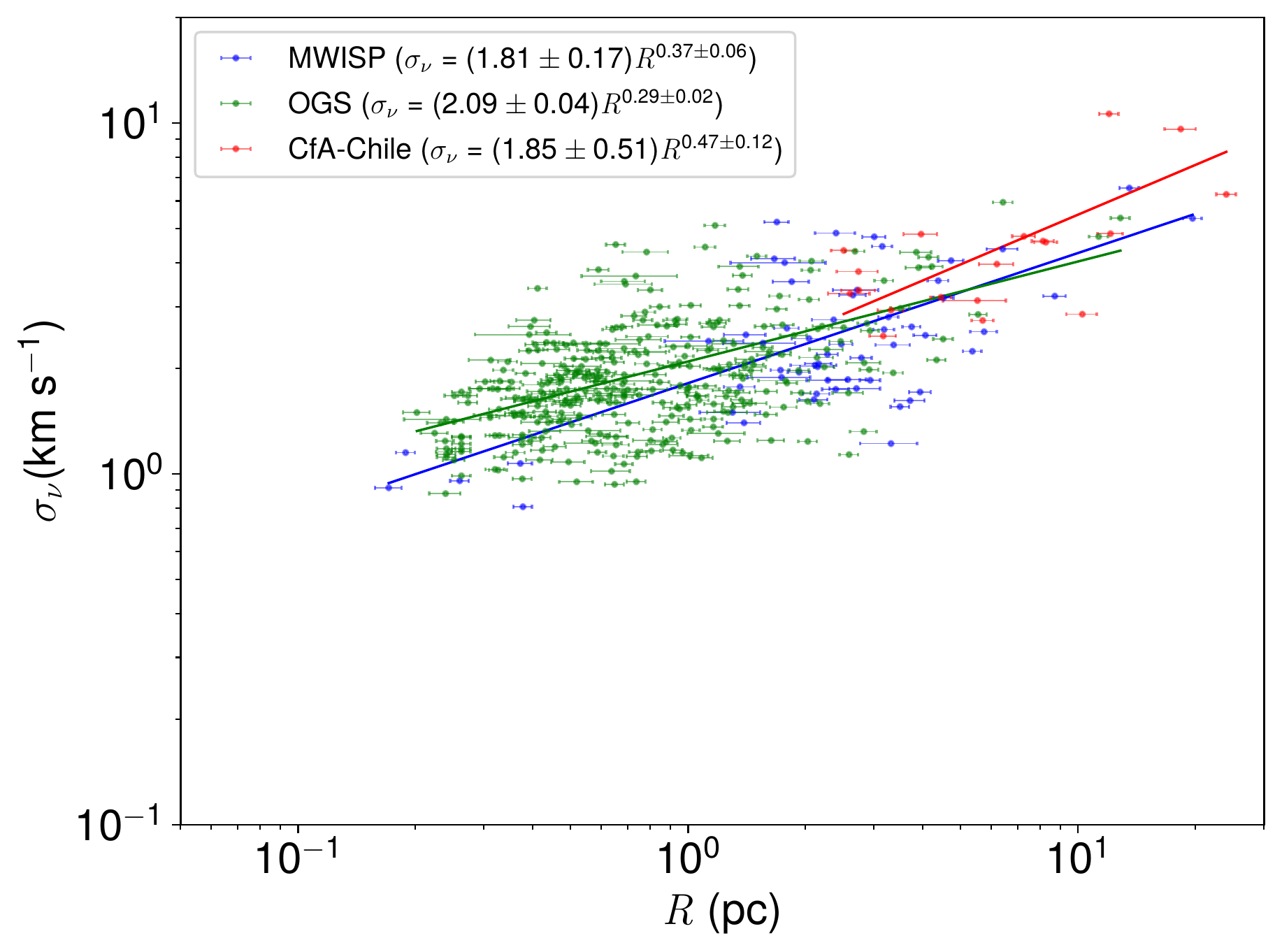}{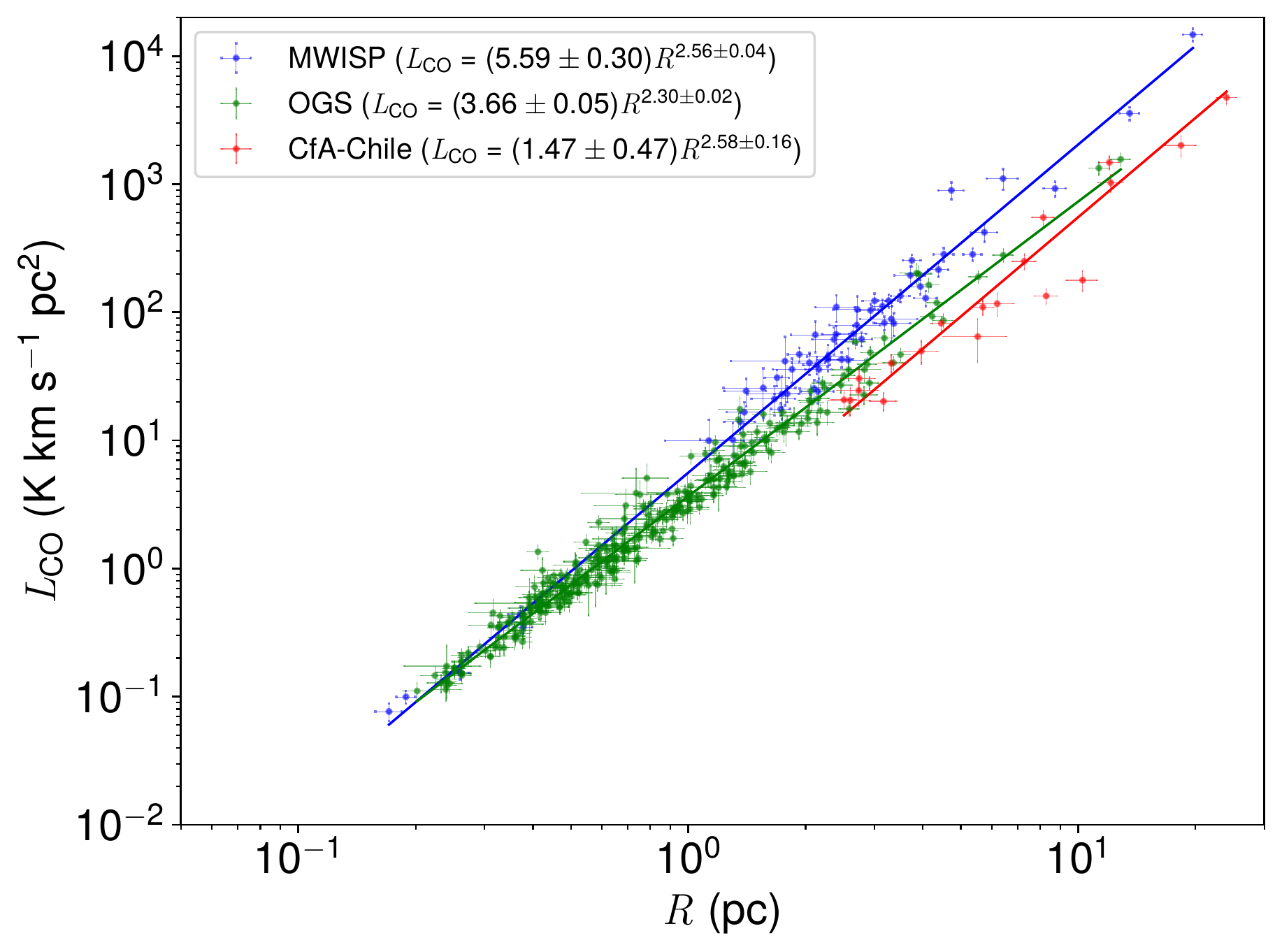}
\caption{Scaling relations of $\sigma_{v}$-$R$ (left) and and $L_{\rm CO}$-$R$ (right). Results of three CO surveys,  MWISP (blue), OGS (green), and CfA-Chile (red), are color-coded. For the convenience of comparison, same axis ranges are used  with Figure \ref{fig:larsonAlgo}. The cutoff of data cubes is 2$\sigma$, and only errors of distances are considered, including the 5\% systematic error \citep{2021ApJ...922....8Y}. \label{fig:larsonThreeSurvey}} 
\end{figure}


\subsection{Giant Molecular Clouds}

Comparisons of observations indicate that the concept of GMCs needs to consider the uncertainty caused by observations. We have already seen that coherent structures in the CfA-Chile survey are divided into many smaller ones in the MWISP survey and the OGS, and it is also possible that large molecular clouds seen by the MWISP survey are actually assembles of isolated molecular clouds. 

This interplay between the angular resolution and sensitivity causes difficulties in identifying molecular clouds in PPV space. High angular resolution alleviates the overlapping of molecular clouds, and high sensitivity, however, reveals the diffuse emission in the vicinity of molecular clouds, which tends to connect molecular clouds in PPV space.

In order to refer to the GMCs as a concept independent of observation parameters, we suggest to define GMCs as clusters or aggregates of molecular clouds. In other words, GMCs stand for a number of molecular clouds towards particular regions. In the framework of SCIMES, \citet{2015MNRAS.454.2067C} proposed to define GMCs as clusters derived from certain kinds of criteria. In the context of SCIMES, molecular clouds are essentially clusters of dendrogram structures, and GMCs are still single molecular clouds. In this work, however, GMCs are not single DBSCAN molecular clouds, but molecular cloud samples.


It would be interesting to compare power-law distributions of the angular area and flux toward GMCs. As shown in Figures \ref{fig:W3578area} and \ref{fig:W3578flux}, distributions of W3, W5, and NGC 7538 are similar. However, variations of GMC distributions can only be revealed by examining a large number of GMCs.

As an analogy, the relationship between molecular clouds and GMCs resembles that between stars and star clusters. Single stars in clusters, such as low mass and binary stars, can only be resolved out under higher sensitivity and angular resolution. This situation is exactly the case for GMCs, and identifying single molecular clouds in GMCs also requires high angular resolution and sensitivity.






\subsection{Larson Relations}

In this section, we examine scaling relations of molecular clouds. \citet{1981MNRAS.194..809L} proposed three scaling relations, regarding turbulence \citep[e.g.,][]{2004ApJ...615L..45H}, virial equilibrium \citep[e.g.,][]{1987ApJ...319..730S,2018MNRAS.477.2220T,2019MNRAS.490.2648B}, and surface densities \citep[e.g.,][]{2010ApJ...716..433K,2020ApJ...898....3L} of molecular clouds.

Providing a thought investigation of Larson relations is beyond the scope of this work, and alternatively, we compare relations between the CO luminosity ($L_{\rm CO}$), the velocity dispersion ($\sigma_{v}$), and the radius ($R$), to see the effects of algorithms and observations. $L_{\rm CO}$ is calculated with the formula given by \citet{2005ARA&A..43..677S}. The velocity dispersion is defined as the equivalent line width of the molecular cloud average spectrum \citep[see][]{2020ApJ...898...80Y}, and the radius is estimated according to the area of molecular clouds, with the beam dilution effect removed.

Figure \ref{fig:larsonAlgo} demonstrates the effect of algorithms on scaling relations, and only those molecular clouds with accurate distances measured by \citet{2021ApJ...922....8Y} are shown.  All molecular clouds that have parts in the PPV range of molecular clouds identified in \citet{2021ApJ...922....8Y} are assigned with the same distance measured. Given the low cutoff (2$\sigma$) used by \citet{2021ApJ...922....8Y}, the distance match is unique.

Effects of algorithms and observations on scaling relations are different. With the same data set but different algorithms, variations of $\sigma_{v}$-$R$ indexes can be significant, but the $L_{\rm CO}$-$R$ relation is quite consistent. However, both scaling relations show evident discrepancies between three CO surveys, as shown in Figure \ref{fig:larsonThreeSurvey}.

In addition to the equivalent line width, the velocity dispersion can also be defined as the second momentum in velocity space \citep[e.g.,][]{2008ApJ...679.1338R}. We also examined the $\sigma_{v}$-$R$ relation under this definition, and found that the relation is approximately the same with those two definitions of velocity dispersion.

The $\sigma_{v}$-$R$  index was first found to be 0.38 by \citet{1981MNRAS.194..809L}, but later revised to be 0.5 \citep[e.g.,][]{2004ApJ...615L..45H}. Interestingly, we found both values in comparisons of algorithms and CO surveys. To understand this behavior, further investigations are needed, involving samples with larger dynamic ranges, multiple line tracers, and different regions.


The $L_{\rm CO}$-$R$ index is roughly constant, about 2.5.  Due to the proportionality between mass and $L_{\rm CO}$, the $L_{\rm CO}$-$R$ index roughly represent the index of the mass-size relation. The conversion factor from $L_{\rm CO}$ to mass is about 4.36 \msun\ $(\rm K\ km\ s^{-1}\ pc^2)^{-1}$, based on a \cofs-to-$\rm H_2$ mass conversion factor of $X=2.0\times10^{20}\ \rm cm^{-2}\ (K\ km\ s^{-1} )^{-1}$. The power-law index of the mass-size relation is about 2 based on extinction maps \citep[e.g.,][]{2010A&A...519L...7L}, indicating a constant surface density of molecular clouds \citep[e.g.,][]{2020ApJ...898....3L}. CO studies, however, suggested larger values, about 2.2-2.3 \citep[e.g.,][]{2010ApJ...723..492R,2017ApJ...834...57M}, and we found even larger indexes. \citet{2019MNRAS.490.2648B} attributed this discrepancy to velocity crowding in PPV space. Given the inconsistency between three CO surveys (see Figure \ref{fig:larsonThreeSurvey}), larger molecular cloud samples with accurate distances are needed to clarify this discrepancy.


Discrepancies in scaling relations indicate that collectively using molecular cloud catalogs obtained at different angular resolution and sensitivity requires caution. Both algorithms and observations change the boundary of molecular clouds, but observations impose two additional effects. First, different angular resolution corresponds to different beam filling factors \citep{2021ApJ...910..109Y}, altering the brightness of molecular clouds. Secondly, sensitivity determines the completeness of molecular cloud catalog, causing biases in statistics.


\section{Summary}
\label{sec:summary}
 
We have examined the variation of molecular cloud samples with respect to sensitivity, angular and velocity resolution, and algorithms. This investigation is based on high quality observations of the MWISP CO survey, and as a verification, we also include the OGS and the CfA-Chile survey. Conclusions are summarized below.


\begin{enumerate}
  
  \item Molecular cloud catalogs are not unique, depending on  angular/velocity resolution, sensitivity, and algorithms.  Low angular/velocity resolution   merge molecular clouds in PPV space, while at low sensitivity, diffuse emission of molecular clouds and small faint molecular clouds are likely to be missed. Catalogs derived with DBSCAN and HDBSCAN are similar, while SCIMES catalogs show slightly larger fluctuations.

\item Different algorithms produce distinguish molecular cloud catalogs, and the $\sigma_{v}$-$R$ relation shows variations with algorithms. The $L_{\rm CO}$-$R$ relation, however, is only slightly altered by algorithms. Three CO surveys, the MWISP, OGS, and the CfA-Chile, observe quite different scaling relations, indicating that observational effects are more fundamental.
  

  \item Using the portion of molecular clouds that are single along lines of sight as a measurement of the velocity crowding, the value is about 8.8\% for the MWISP survey in the second Galactic quadrant, meaning that the velocity crowding effect is significant for \cofs\ clouds even toward the outer Galaxy.
  

   \item The MWISP CO survey suggests soft or smoothing molecular cloud boundaries, indicating a picture of turbulent density fluctuations in the formation of molecular clouds.

  \item Due to the increased sensitivity and resolving power, GMCs are resolved into multiple independent clouds in PPV space, which should be a new picture of GMCs.  In other words, GMCs are collections or aggregates of molecular clouds, i.e., GMCs themselves are molecular cloud samples. The angular area and flux distributions of GMCs are also power laws.


 
\end{enumerate}

\begin{acknowledgments}

This research made use of the data from the Milky Way Imaging Scroll Painting (MWISP) project, which is a multi-line survey in \cofs/\coss/\cots\ along the northern Galactic plane with PMO-13.7m telescope. We are grateful to all the members of the MWISP working group, particularly Min Wang, Jixian Sun,  Dengrong Lu, and the staff members at PMO-13.7m telescope, for their long-term support. We acknowledge Xuepeng Chen, Lixia Yuan, Yuehui Ma, Fujun Du, and Min Fang for their helpful discussions. This work was supported by National Natural Science Foundation of China through grants 12041305 \& 12003071. MWISP was sponsored by National Key R\&D Program of China with grant 2017YFA0402701 and by CAS Key Research Program of Frontier Sciences with grant QYZDJ-SSW-SLH047. 
\end{acknowledgments} 
%

\vspace{5mm}
\facilities{PMO-13.7m}


\software{astropy \citep{2013A&A...558A..33A},   SciPy \citep{2020SciPy-NMeth}, DBSCAN \citep{citeDBSCAN}, HDBSCAN \citep{citeHDBSCAN}, SCIMES \citep{2015MNRAS.454.2067C}.
          }

 \bibliographystyle{aasjournal}
 \bibliography{refMCsample}

\begin{thebibliography}{}
\expandafter\ifx\csname natexlab\endcsname\relax\def\natexlab#1{#1}\fi
\providecommand{\url}[1]{\href{#1}{#1}}
\providecommand{\dodoi}[1]{doi:~\href{http://doi.org/#1}{\nolinkurl{#1}}}
\providecommand{\doeprint}[1]{\href{http://ascl.net/#1}{\nolinkurl{http://ascl.net/#1}}}
\providecommand{\doarXiv}[1]{\href{https://arxiv.org/abs/#1}{\nolinkurl{https://arxiv.org/abs/#1}}}

\bibitem[{{Acharyya} {et~al.}(2020){Acharyya}, {Krumholz}, {Federrath},
  {Kewley}, {Goldbaum}, \& {Sharp}}]{2020MNRAS.495.3819A}
{Acharyya}, A., {Krumholz}, M.~R., {Federrath}, C., {et~al.} 2020, \mnras, 495,
  3819, \dodoi{10.1093/mnras/staa1100}

\bibitem[{{Arzoumanian} {et~al.}(2019){Arzoumanian}, {Andr{\'e}},
  {K{\"o}nyves}, {Palmeirim}, {Roy}, {Schneider}, {Benedettini}, {Didelon}, {Di
  Francesco}, {Kirk}, \& {Ladjelate}}]{2019A&A...621A..42A}
{Arzoumanian}, D., {Andr{\'e}}, P., {K{\"o}nyves}, V., {et~al.} 2019, \aap,
  621, A42, \dodoi{10.1051/0004-6361/201832725}

\bibitem[{{Astropy Collaboration} {et~al.}(2013){Astropy Collaboration},
  {Robitaille}, {Tollerud}, {Greenfield}, {Droettboom}, {Bray}, {Aldcroft},
  {Davis}, {Ginsburg}, {Price-Whelan}, {Kerzendorf}, {Conley}, {Crighton},
  {Barbary}, {Muna}, {Ferguson}, {Grollier}, {Parikh}, {Nair}, {Unther},
  {Deil}, {Woillez}, {Conseil}, {Kramer}, {Turner}, {Singer}, {Fox}, {Weaver},
  {Zabalza}, {Edwards}, {Azalee Bostroem}, {Burke}, {Casey}, {Crawford},
  {Dencheva}, {Ely}, {Jenness}, {Labrie}, {Lim}, {Pierfederici}, {Pontzen},
  {Ptak}, {Refsdal}, {Servillat}, \& {Streicher}}]{2013A&A...558A..33A}
{Astropy Collaboration}, {Robitaille}, T.~P., {Tollerud}, E.~J., {et~al.} 2013,
  \aap, 558, A33, \dodoi{10.1051/0004-6361/201322068}

\bibitem[{{Ballesteros-Paredes} {et~al.}(2019){Ballesteros-Paredes},
  {Rom{\'a}n-Z{\'u}{\~n}iga}, {Salom{\'e}}, {Zamora-Avil{\'e}s}, \&
  {Jim{\'e}nez-Donaire}}]{2019MNRAS.490.2648B}
{Ballesteros-Paredes}, J., {Rom{\'a}n-Z{\'u}{\~n}iga}, C., {Salom{\'e}}, Q.,
  {Zamora-Avil{\'e}s}, M., \& {Jim{\'e}nez-Donaire}, M.~J. 2019, \mnras, 490,
  2648, \dodoi{10.1093/mnras/stz2575}

\bibitem[{{Ballesteros-Paredes} {et~al.}(1999){Ballesteros-Paredes},
  {V{\'a}zquez-Semadeni}, \& {Scalo}}]{1999ApJ...515..286B}
{Ballesteros-Paredes}, J., {V{\'a}zquez-Semadeni}, E., \& {Scalo}, J. 1999,
  \apj, 515, 286, \dodoi{10.1086/307007}

\bibitem[{{Bialy} {et~al.}(2021){Bialy}, {Zucker}, {Goodman}, {Foley}, {Alves},
  {Semenov}, {Benjamin}, {Leike}, \& {En{\ss}lin}}]{2021ApJ...919L...5B}
{Bialy}, S., {Zucker}, C., {Goodman}, A., {et~al.} 2021, \apjl, 919, L5,
  \dodoi{10.3847/2041-8213/ac1f95}

\bibitem[{{Bieging} \& {Peters}(2011)}]{2011ApJS..196...18B}
{Bieging}, J.~H., \& {Peters}, W.~L. 2011, \apjs, 196, 18,
  \dodoi{10.1088/0067-0049/196/2/18}

\bibitem[{{Bigiel} {et~al.}(2008){Bigiel}, {Leroy}, {Walter}, {Brinks}, {de
  Blok}, {Madore}, \& {Thornley}}]{2008AJ....136.2846B}
{Bigiel}, F., {Leroy}, A., {Walter}, F., {et~al.} 2008, \aj, 136, 2846,
  \dodoi{10.1088/0004-6256/136/6/2846}

\bibitem[{{Blitz} \& {Thaddeus}(1980)}]{1980ApJ...241..676B}
{Blitz}, L., \& {Thaddeus}, P. 1980, \apj, 241, 676, \dodoi{10.1086/158379}

\bibitem[{{Bolatto} {et~al.}(2013){Bolatto}, {Wolfire}, \&
  {Leroy}}]{2013ARA&A..51..207B}
{Bolatto}, A.~D., {Wolfire}, M., \& {Leroy}, A.~K. 2013, \araa, 51, 207,
  \dodoi{10.1146/annurev-astro-082812-140944}

\bibitem[{{Bot} {et~al.}(2007){Bot}, {Boulanger}, {Rubio}, \&
  {Rantakyro}}]{2007A&A...471..103B}
{Bot}, C., {Boulanger}, F., {Rubio}, M., \& {Rantakyro}, F. 2007, \aap, 471,
  103, \dodoi{10.1051/0004-6361:20066612}

\bibitem[{{Brunt} {et~al.}(2003){Brunt}, {Kerton}, \&
  {Pomerleau}}]{2003ApJS..144...47B}
{Brunt}, C.~M., {Kerton}, C.~R., \& {Pomerleau}, C. 2003, \apjs, 144, 47,
  \dodoi{10.1086/344245}

\bibitem[{{Campbell}(1984)}]{1984ApJ...282L..27C}
{Campbell}, B. 1984, \apjl, 282, L27, \dodoi{10.1086/184297}

\bibitem[{Campello {et~al.}(2013)Campello, Moulavi, \& Sander}]{citeHDBSCAN}
Campello, R. J. G.~B., Moulavi, D., \& Sander, J. 2013, in Advances in
  Knowledge Discovery and Data Mining, ed. J.~Pei, V.~S. Tseng, L.~Cao,
  H.~Motoda, \& G.~Xu (Berlin, Heidelberg: Springer Berlin Heidelberg),
  160--172, \dodoi{10.1007/978-3-642-37456-2_14}

\bibitem[{{Chevance} {et~al.}(2020){Chevance}, {Kruijssen}, {Vazquez-Semadeni},
  {Nakamura}, {Klessen}, {Ballesteros-Paredes}, {Inutsuka}, {Adamo}, \&
  {Hennebelle}}]{2020SSRv..216...50C}
{Chevance}, M., {Kruijssen}, J.~M.~D., {Vazquez-Semadeni}, E., {et~al.} 2020,
  \ssr, 216, 50, \dodoi{10.1007/s11214-020-00674-x}

\bibitem[{{Clarke} {et~al.}(2018){Clarke}, {Whitworth}, {Spowage},
  {Duarte-Cabral}, {Suri}, {Jaffa}, {Walch}, \& {Clark}}]{2018MNRAS.479.1722C}
{Clarke}, S.~D., {Whitworth}, A.~P., {Spowage}, R.~L., {et~al.} 2018, \mnras,
  479, 1722, \dodoi{10.1093/mnras/sty1675}

\bibitem[{{Colombo} {et~al.}(2015){Colombo}, {Rosolowsky}, {Ginsburg},
  {Duarte-Cabral}, \& {Hughes}}]{2015MNRAS.454.2067C}
{Colombo}, D., {Rosolowsky}, E., {Ginsburg}, A., {Duarte-Cabral}, A., \&
  {Hughes}, A. 2015, \mnras, 454, 2067, \dodoi{10.1093/mnras/stv2063}

\bibitem[{{Colombo} {et~al.}(2019){Colombo}, {Rosolowsky}, {Duarte-Cabral},
  {Ginsburg}, {Glenn}, {Zetterlund}, {Hernand ez}, {Dempsey}, \&
  {Currie}}]{2019MNRAS.483.4291C}
{Colombo}, D., {Rosolowsky}, E., {Duarte-Cabral}, A., {et~al.} 2019, \mnras,
  483, 4291, \dodoi{10.1093/mnras/sty3283}

\bibitem[{{Cong}(1977)}]{1977PhDT.......123C}
{Cong}, H.~I.~L. 1977, PhD thesis, Columbia University

\bibitem[{{Dame} {et~al.}(2001){Dame}, {Hartmann}, \&
  {Thaddeus}}]{2001ApJ...547..792D}
{Dame}, T.~M., {Hartmann}, D., \& {Thaddeus}, P. 2001, \apj, 547, 792,
  \dodoi{10.1086/318388}

\bibitem[{{Deharveng} {et~al.}(2012){Deharveng}, {Zavagno}, {Anderson},
  {Motte}, {Abergel}, {Andr{\'e}}, {Bontemps}, {Leleu}, {Roussel}, \&
  {Russeil}}]{2012A&A...546A..74D}
{Deharveng}, L., {Zavagno}, A., {Anderson}, L.~D., {et~al.} 2012, \aap, 546,
  A74, \dodoi{10.1051/0004-6361/201219131}

\bibitem[{{Dempsey} {et~al.}(2013){Dempsey}, {Thomas}, \&
  {Currie}}]{2013ApJS..209....8D}
{Dempsey}, J.~T., {Thomas}, H.~S., \& {Currie}, M.~J. 2013, \apjs, 209, 8,
  \dodoi{10.1088/0067-0049/209/1/8}

\bibitem[{{den Brok} {et~al.}(2021){den Brok}, {Chatzigiannakis}, {Bigiel},
  {Puschnig}, {Barnes}, {Leroy}, {Jim{\'e}nez-Donaire}, {Usero}, {Schinnerer},
  {Rosolowsky}, {Faesi}, {Grasha}, {Hughes}, {Kruijssen}, {Liu}, {Neumann},
  {Pety}, {Querejeta}, {Saito}, {Schruba}, \& {Stuber}}]{2021MNRAS.504.3221D}
{den Brok}, J.~S., {Chatzigiannakis}, D., {Bigiel}, F., {et~al.} 2021, \mnras,
  504, 3221, \dodoi{10.1093/mnras/stab859}

\bibitem[{{Dickel} {et~al.}(1981){Dickel}, {Dickel}, \&
  {Wilson}}]{1981ApJ...250L..43D}
{Dickel}, H.~R., {Dickel}, J.~R., \& {Wilson}, W.~J. 1981, \apjl, 250, L43,
  \dodoi{10.1086/183671}

\bibitem[{{Dickel} {et~al.}(1980){Dickel}, {Dickel}, {Wilson}, \&
  {Werner}}]{1980ApJ...237..711D}
{Dickel}, H.~R., {Dickel}, J.~R., {Wilson}, W.~J., \& {Werner}, M.~W. 1980,
  \apj, 237, 711, \dodoi{10.1086/157919}

\bibitem[{{Digel} {et~al.}(1996){Digel}, {Lyder}, {Philbrick}, {Puche}, \&
  {Thaddeus}}]{1996ApJ...458..561D}
{Digel}, S.~W., {Lyder}, D.~A., {Philbrick}, A.~J., {Puche}, D., \& {Thaddeus},
  P. 1996, \apj, 458, 561, \dodoi{10.1086/176839}

\bibitem[{{Dobbs} \& {Pringle}(2013)}]{2013MNRAS.432..653D}
{Dobbs}, C.~L., \& {Pringle}, J.~E. 2013, \mnras, 432, 653,
  \dodoi{10.1093/mnras/stt508}

\bibitem[{{Druard} {et~al.}(2014){Druard}, {Braine}, {Schuster}, {Schneider},
  {Gratier}, {Bontemps}, {Boquien}, {Combes}, {Corbelli}, {Henkel}, {Herpin},
  {Kramer}, {van der Tak}, \& {van der Werf}}]{2014A&A...567A.118D}
{Druard}, C., {Braine}, J., {Schuster}, K.~F., {et~al.} 2014, \aap, 567, A118,
  \dodoi{10.1051/0004-6361/201423682}

\bibitem[{{Du} {et~al.}(2017){Du}, {Xu}, {Yang}, \&
  {Sun}}]{2017ApJS..229...24D}
{Du}, X., {Xu}, Y., {Yang}, J., \& {Sun}, Y. 2017, \apjs, 229, 24,
  \dodoi{10.3847/1538-4365/aa5d9d}

\bibitem[{{Duarte-Cabral} \& {Dobbs}(2016)}]{2016MNRAS.458.3667D}
{Duarte-Cabral}, A., \& {Dobbs}, C.~L. 2016, \mnras, 458, 3667,
  \dodoi{10.1093/mnras/stw469}

\bibitem[{{Duarte-Cabral} \& {Dobbs}(2017)}]{2017MNRAS.470.4261D}
---. 2017, \mnras, 470, 4261, \dodoi{10.1093/mnras/stx1524}

\bibitem[{{Duarte-Cabral} {et~al.}(2021){Duarte-Cabral}, {Colombo}, {Urquhart},
  {Ginsburg}, {Russeil}, {Schuller}, {Anderson}, {Barnes}, {Beltr{\'a}n},
  {Beuther}, {Bontemps}, {Bronfman}, {Csengeri}, {Dobbs}, {Eden}, {Giannetti},
  {Kauffmann}, {Mattern}, {Medina}, {Menten}, {Lee}, {Pettitt}, {Riener},
  {Rigby}, {Traficante}, {Veena}, {Wienen}, {Wyrowski}, {Agurto}, {Azagra},
  {Cesaroni}, {Finger}, {Gonzalez}, {Henning}, {Hernandez}, {Kainulainen},
  {Leurini}, {Lopez}, {Mac-Auliffe}, {Mazumdar}, {Molinari}, {Motte}, {Muller},
  {Nguyen-Luong}, {Parra}, {Perez-Beaupuits}, {Montenegro-Montes}, {Moore},
  {Ragan}, {S{\'a}nchez-Monge}, {Sanna}, {Schilke}, {Schisano}, {Schneider},
  {Suri}, {Testi}, {Torstensson}, {Venegas}, {Wang}, \&
  {Zavagno}}]{2021MNRAS.500.3027D}
{Duarte-Cabral}, A., {Colombo}, D., {Urquhart}, J.~S., {et~al.} 2021, \mnras,
  500, 3027, \dodoi{10.1093/mnras/staa2480}

\bibitem[{Ester {et~al.}(1996)Ester, Kriegel, Sander, \& Xu}]{citeDBSCAN}
Ester, M., Kriegel, H.-P., Sander, J., \& Xu, X. 1996, in Proceedings of the
  Second International Conference on Knowledge Discovery and Data Mining,
  KDD'96 (AAAI Press), 226--231.
\newblock \url{http://dl.acm.org/citation.cfm?id=3001460.3001507}

\bibitem[{{Evans} {et~al.}(2021){Evans}, {Heyer}, {Miville-Desch{\^e}nes},
  {Nguyen-Luong}, \& {Merello}}]{2021ApJ...920..126E}
{Evans}, Neal~J., I., {Heyer}, M., {Miville-Desch{\^e}nes}, M.-A.,
  {Nguyen-Luong}, Q., \& {Merello}, M. 2021, \apj, 920, 126,
  \dodoi{10.3847/1538-4357/ac1425}

\bibitem[{{Faesi} {et~al.}(2018){Faesi}, {Lada}, \&
  {Forbrich}}]{2018ApJ...857...19F}
{Faesi}, C.~M., {Lada}, C.~J., \& {Forbrich}, J. 2018, \apj, 857, 19,
  \dodoi{10.3847/1538-4357/aaad60}

\bibitem[{{Gao} {et~al.}(2007){Gao}, {Carilli}, {Solomon}, \& {Vanden
  Bout}}]{2007ApJ...660L..93G}
{Gao}, Y., {Carilli}, C.~L., {Solomon}, P.~M., \& {Vanden Bout}, P.~A. 2007,
  \apjl, 660, L93, \dodoi{10.1086/518244}

\bibitem[{{Gao} \& {Solomon}(2004)}]{2004ApJ...606..271G}
{Gao}, Y., \& {Solomon}, P.~M. 2004, \apj, 606, 271, \dodoi{10.1086/382999}

\bibitem[{{Ginsburg} {et~al.}(2011){Ginsburg}, {Bally}, \&
  {Williams}}]{2011MNRAS.418.2121G}
{Ginsburg}, A., {Bally}, J., \& {Williams}, J.~P. 2011, \mnras, 418, 2121,
  \dodoi{10.1111/j.1365-2966.2011.19279.x}

\bibitem[{{Goldsmith} {et~al.}(2008){Goldsmith}, {Heyer}, {Narayanan}, {Snell},
  {Li}, \& {Brunt}}]{2008ApJ...680..428G}
{Goldsmith}, P.~F., {Heyer}, M., {Narayanan}, G., {et~al.} 2008, \apj, 680,
  428, \dodoi{10.1086/587166}

\bibitem[{{Guszejnov} {et~al.}(2020){Guszejnov}, {Grudi{\'c}}, {Offner},
  {Boylan-Kolchin}, {Faucher-Gigu{\`e}re}, {Wetzel}, {Benincasa}, \&
  {Loebman}}]{2020MNRAS.492..488G}
{Guszejnov}, D., {Grudi{\'c}}, M.~Y., {Offner}, S. S.~R., {et~al.} 2020,
  \mnras, 492, 488, \dodoi{10.1093/mnras/stz3527}

\bibitem[{{Heyer} \& {Dame}(2015)}]{2015ARA&A..53..583H}
{Heyer}, M., \& {Dame}, T.~M. 2015, \araa, 53, 583,
  \dodoi{10.1146/annurev-astro-082214-122324}

\bibitem[{{Heyer} {et~al.}(1998){Heyer}, {Brunt}, {Snell}, {Howe}, {Schloerb},
  \& {Carpenter}}]{1998ApJS..115..241H}
{Heyer}, M.~H., {Brunt}, C., {Snell}, R.~L., {et~al.} 1998, \apjs, 115, 241,
  \dodoi{10.1086/313086}

\bibitem[{{Heyer} \& {Brunt}(2004)}]{2004ApJ...615L..45H}
{Heyer}, M.~H., \& {Brunt}, C.~M. 2004, \apjl, 615, L45, \dodoi{10.1086/425978}

\bibitem[{{Heyer} {et~al.}(2001){Heyer}, {Carpenter}, \&
  {Snell}}]{2001ApJ...551..852H}
{Heyer}, M.~H., {Carpenter}, J.~M., \& {Snell}, R.~L. 2001, \apj, 551, 852,
  \dodoi{10.1086/320218}

\bibitem[{{Heyer} \& {Terebey}(1998)}]{1998ApJ...502..265H}
{Heyer}, M.~H., \& {Terebey}, S. 1998, \apj, 502, 265, \dodoi{10.1086/305881}

\bibitem[{{Hughes} {et~al.}(2013){Hughes}, {Meidt}, {Colombo}, {Schinnerer},
  {Pety}, {Leroy}, {Dobbs}, {Garc{\'\i}a-Burillo}, {Thompson}, {Dumas},
  {Schuster}, \& {Kramer}}]{2013ApJ...779...46H}
{Hughes}, A., {Meidt}, S.~E., {Colombo}, D., {et~al.} 2013, \apj, 779, 46,
  \dodoi{10.1088/0004-637X/779/1/46}

\bibitem[{{Jackson} {et~al.}(2006){Jackson}, {Rathborne}, {Shah}, {Simon},
  {Bania}, {Clemens}, {Chambers}, {Johnson}, {Dormody}, {Lavoie}, \&
  {Heyer}}]{2006ApJS..163..145J}
{Jackson}, J.~M., {Rathborne}, J.~M., {Shah}, R.~Y., {et~al.} 2006, \apjs, 163,
  145, \dodoi{10.1086/500091}

\bibitem[{{Karr} \& {Martin}(2003)}]{2003ApJ...595..900K}
{Karr}, J.~L., \& {Martin}, P.~G. 2003, \apj, 595, 900, \dodoi{10.1086/376590}

\bibitem[{{Kauffmann} {et~al.}(2017){Kauffmann}, {Goldsmith}, {Melnick},
  {Tolls}, {Guzman}, \& {Menten}}]{2017A&A...605L...5K}
{Kauffmann}, J., {Goldsmith}, P.~F., {Melnick}, G., {et~al.} 2017, \aap, 605,
  L5, \dodoi{10.1051/0004-6361/201731123}

\bibitem[{{Kauffmann} {et~al.}(2010){Kauffmann}, {Pillai}, {Shetty}, {Myers},
  \& {Goodman}}]{2010ApJ...716..433K}
{Kauffmann}, J., {Pillai}, T., {Shetty}, R., {Myers}, P.~C., \& {Goodman},
  A.~A. 2010, \apj, 716, 433, \dodoi{10.1088/0004-637X/716/1/433}

\bibitem[{{Kennicutt}(1989)}]{1989ApJ...344..685K}
{Kennicutt}, Robert~C., J. 1989, \apj, 344, 685, \dodoi{10.1086/167834}

\bibitem[{{Kennicutt} \& {Evans}(2012)}]{2012ARA&A..50..531K}
{Kennicutt}, R.~C., \& {Evans}, N.~J. 2012, \araa, 50, 531,
  \dodoi{10.1146/annurev-astro-081811-125610}

\bibitem[{{Kodaira} {et~al.}(1977){Kodaira}, {Ishii}, {Nakamura}, {Inatani},
  {Nagane}, {Tojo}, \& {Sato}}]{1977PASJ...29...53K}
{Kodaira}, S., {Ishii}, K., {Nakamura}, T., {et~al.} 1977, \pasj, 29, 53

\bibitem[{{Koenig} {et~al.}(2008){Koenig}, {Allen}, {Gutermuth}, {Hora},
  {Brunt}, \& {Muzerolle}}]{2008ApJ...688.1142K}
{Koenig}, X.~P., {Allen}, L.~E., {Gutermuth}, R.~A., {et~al.} 2008, \apj, 688,
  1142, \dodoi{10.1086/592322}

\bibitem[{{K{\"o}nyves} {et~al.}(2015){K{\"o}nyves}, {Andr{\'e}},
  {Men'shchikov}, {Palmeirim}, {Arzoumanian}, {Schneider}, {Roy}, {Didelon},
  {Maury}, {Shimajiri}, {Di Francesco}, {Bontemps}, {Peretto}, {Benedettini},
  {Bernard}, {Elia}, {Griffin}, {Hill}, {Kirk}, {Ladjelate}, {Marsh}, {Martin},
  {Motte}, {Nguy{\^e}n Luong}, {Pezzuto}, {Roussel}, {Rygl}, {Sadavoy},
  {Schisano}, {Spinoglio}, {Ward-Thompson}, \& {White}}]{2015A&A...584A..91K}
{K{\"o}nyves}, V., {Andr{\'e}}, P., {Men'shchikov}, A., {et~al.} 2015, \aap,
  584, A91, \dodoi{10.1051/0004-6361/201525861}

\bibitem[{{Kutner} \& {Tucker}(1975)}]{1975ApJ...199...79K}
{Kutner}, M.~L., \& {Tucker}, K.~D. 1975, \apj, 199, 79, \dodoi{10.1086/153666}

\bibitem[{{Kutner} {et~al.}(1977){Kutner}, {Tucker}, {Chin}, \&
  {Thaddeus}}]{1977ApJ...215..521K}
{Kutner}, M.~L., {Tucker}, K.~D., {Chin}, G., \& {Thaddeus}, P. 1977, \apj,
  215, 521, \dodoi{10.1086/155384}

\bibitem[{{Kutner} \& {Ulich}(1981)}]{1981ApJ...250..341K}
{Kutner}, M.~L., \& {Ulich}, B.~L. 1981, \apj, 250, 341, \dodoi{10.1086/159380}

\bibitem[{{Lada} \& {Dame}(2020)}]{2020ApJ...898....3L}
{Lada}, C.~J., \& {Dame}, T.~M. 2020, \apj, 898, 3,
  \dodoi{10.3847/1538-4357/ab9bfb}

\bibitem[{{Larson}(1981)}]{1981MNRAS.194..809L}
{Larson}, R.~B. 1981, \mnras, 194, 809, \dodoi{10.1093/mnras/194.4.809}

\bibitem[{{Leroy} {et~al.}(2015){Leroy}, {Walter}, {Martini}, {Roussel},
  {Sandstrom}, {Ott}, {Weiss}, {Bolatto}, {Schuster}, \&
  {Dessauges-Zavadsky}}]{2015ApJ...814...83L}
{Leroy}, A.~K., {Walter}, F., {Martini}, P., {et~al.} 2015, \apj, 814, 83,
  \dodoi{10.1088/0004-637X/814/2/83}

\bibitem[{{Leung} \& {Thaddeus}(1992)}]{1992ApJS...81..267L}
{Leung}, H.~O., \& {Thaddeus}, P. 1992, \apjs, 81, 267, \dodoi{10.1086/191693}

\bibitem[{{Li} {et~al.}(2019){Li}, {Xu}, {Sun}, {Yan}, {Ma}, \&
  {Yang}}]{2019ApJS..242...19L}
{Li}, Y., {Xu}, Y., {Sun}, Y., {et~al.} 2019, \apjs, 242, 19,
  \dodoi{10.3847/1538-4365/ab1e55}

\bibitem[{{Lim} {et~al.}(2014){Lim}, {Sung}, {Kim}, {Bessell}, \&
  {Karimov}}]{2014MNRAS.438.1451L}
{Lim}, B., {Sung}, H., {Kim}, J.~S., {Bessell}, M.~S., \& {Karimov}, R. 2014,
  \mnras, 438, 1451, \dodoi{10.1093/mnras/stt2283}

\bibitem[{{Liszt} {et~al.}(1974){Liszt}, {Wilson}, {Penzias}, {Jefferts},
  {Wannier}, \& {Solomon}}]{1974ApJ...190..557L}
{Liszt}, H.~S., {Wilson}, R.~W., {Penzias}, A.~A., {et~al.} 1974, \apj, 190,
  557, \dodoi{10.1086/152910}

\bibitem[{{Lombardi} {et~al.}(2010){Lombardi}, {Alves}, \&
  {Lada}}]{2010A&A...519L...7L}
{Lombardi}, M., {Alves}, J., \& {Lada}, C.~J. 2010, \aap, 519, L7,
  \dodoi{10.1051/0004-6361/201015282}

\bibitem[{{Louvet} {et~al.}(2021){Louvet}, {Hennebelle}, {Men'shchikov},
  {Didelon}, {Ntormousi}, \& {Motte}}]{2021A&A...653A.157L}
{Louvet}, F., {Hennebelle}, P., {Men'shchikov}, A., {et~al.} 2021, \aap, 653,
  A157, \dodoi{10.1051/0004-6361/202040053}

\bibitem[{{Maeda} {et~al.}(2022){Maeda}, {Egusa}, {Ohta}, {Fujimoto}, {Habe},
  \& {Asada}}]{2022ApJ...926...96M}
{Maeda}, F., {Egusa}, F., {Ohta}, K., {et~al.} 2022, \apj, 926, 96,
  \dodoi{10.3847/1538-4357/ac4505}

\bibitem[{{Mazumdar} {et~al.}(2021){Mazumdar}, {Wyrowski}, {Colombo},
  {Urquhart}, {Thompson}, \& {Menten}}]{2021A&A...650A.164M}
{Mazumdar}, P., {Wyrowski}, F., {Colombo}, D., {et~al.} 2021, \aap, 650, A164,
  \dodoi{10.1051/0004-6361/202040205}

\bibitem[{{Megeath} {et~al.}(2008){Megeath}, {Townsley}, {Oey}, \&
  {Tieftrunk}}]{2008hsf1.book..264M}
{Megeath}, S.~T., {Townsley}, L.~K., {Oey}, M.~S., \& {Tieftrunk}, A.~R. 2008,
  {Low and High Mass Star Formation in the W3, W4, and W5 Regions}, ed.
  B.~{Reipurth}, Vol.~4, 264

\bibitem[{{Melchior} \& {Combes}(2016)}]{2016A&A...585A..44M}
{Melchior}, A.-L., \& {Combes}, F. 2016, \aap, 585, A44,
  \dodoi{10.1051/0004-6361/201526257}

\bibitem[{{Minier} {et~al.}(2000){Minier}, {Booth}, \&
  {Conway}}]{2000A&A...362.1093M}
{Minier}, V., {Booth}, R.~S., \& {Conway}, J.~E. 2000, \aap, 362, 1093

\bibitem[{{Miura} {et~al.}(2021){Miura}, {Espada}, {Hirota}, {Henkel},
  {Verley}, {Kobayashi}, {Matsushita}, {Israel}, {Vila-Vilaro},
  {Morokuma-Matsui}, {Ott}, {Vlahakis}, {Peck}, {Aalto}, {Hogerheijde},
  {Neumayer}, {Iono}, {Kohno}, {Takemura}, \& {Komugi}}]{2021MNRAS.504.6198M}
{Miura}, R.~E., {Espada}, D., {Hirota}, A., {et~al.} 2021, \mnras, 504, 6198,
  \dodoi{10.1093/mnras/stab1210}

\bibitem[{{Miville-Desch{\^e}nes} {et~al.}(2017){Miville-Desch{\^e}nes},
  {Murray}, \& {Lee}}]{2017ApJ...834...57M}
{Miville-Desch{\^e}nes}, M.-A., {Murray}, N., \& {Lee}, E.~J. 2017, \apj, 834,
  57, \dodoi{10.3847/1538-4357/834/1/57}

\bibitem[{{Moscadelli} {et~al.}(2009){Moscadelli}, {Reid}, {Menten},
  {Brunthaler}, {Zheng}, \& {Xu}}]{2009ApJ...693..406M}
{Moscadelli}, L., {Reid}, M.~J., {Menten}, K.~M., {et~al.} 2009, \apj, 693,
  406, \dodoi{10.1088/0004-637X/693/1/406}

\bibitem[{{Murphy} \& {Myers}(1985)}]{1985ApJ...298..818M}
{Murphy}, D.~C., \& {Myers}, P.~C. 1985, \apj, 298, 818, \dodoi{10.1086/163663}

\bibitem[{{Oliver} {et~al.}(1996){Oliver}, {Masheder}, \&
  {Thaddeus}}]{1996A&A...315..578O}
{Oliver}, R.~J., {Masheder}, M.~R.~W., \& {Thaddeus}, P. 1996, \aap, 315, 578

\bibitem[{{Pan} {et~al.}(2016){Pan}, {Fujimoto}, {Tasker}, {Rosolowsky},
  {Colombo}, {Benincasa}, \& {Wadsley}}]{2016MNRAS.458.2443P}
{Pan}, H.-A., {Fujimoto}, Y., {Tasker}, E.~J., {et~al.} 2016, \mnras, 458,
  2443, \dodoi{10.1093/mnras/stw478}

\bibitem[{{Panopoulou} {et~al.}(2021){Panopoulou}, {Clark}, {Hacar}, {Heitsch},
  {Kainulainen}, {Ntormousi}, {Seifried}, \& {Smith}}]{2021arXiv211108125P}
{Panopoulou}, G.~V., {Clark}, S.~E., {Hacar}, A., {et~al.} 2021, arXiv
  e-prints, arXiv:2111.08125.
\newblock \doarXiv{2111.08125}

\bibitem[{{Peek} {et~al.}(2022){Peek}, {Tchernyshyov}, \&
  {Miville-Deschenes}}]{2022ApJ...925..201P}
{Peek}, J.~E.~G., {Tchernyshyov}, K., \& {Miville-Deschenes}, M.-A. 2022, \apj,
  925, 201, \dodoi{10.3847/1538-4357/ac3f34}

\bibitem[{{Pichardo} {et~al.}(2000){Pichardo}, {V{\'a}zquez-Semadeni}, {Gazol},
  {Passot}, \& {Ballesteros-Paredes}}]{2000ApJ...532..353P}
{Pichardo}, B., {V{\'a}zquez-Semadeni}, E., {Gazol}, A., {Passot}, T., \&
  {Ballesteros-Paredes}, J. 2000, \apj, 532, 353, \dodoi{10.1086/308546}

\bibitem[{{Pineda} {et~al.}(2009){Pineda}, {Rosolowsky}, \&
  {Goodman}}]{2009ApJ...699L.134P}
{Pineda}, J.~E., {Rosolowsky}, E.~W., \& {Goodman}, A.~A. 2009, \apjl, 699,
  L134, \dodoi{10.1088/0004-637X/699/2/L134}

\bibitem[{{Polychroni} {et~al.}(2012){Polychroni}, {Moore}, \&
  {Allsopp}}]{2012MNRAS.422.2992P}
{Polychroni}, D., {Moore}, T.~J.~T., \& {Allsopp}, J. 2012, \mnras, 422, 2992,
  \dodoi{10.1111/j.1365-2966.2012.20803.x}

\bibitem[{{Rebolledo} {et~al.}(2015){Rebolledo}, {Wong}, {Xue}, {Leroy},
  {Koda}, \& {Donovan Meyer}}]{2015ApJ...808...99R}
{Rebolledo}, D., {Wong}, T., {Xue}, R., {et~al.} 2015, \apj, 808, 99,
  \dodoi{10.1088/0004-637X/808/1/99}

\bibitem[{{Rice} {et~al.}(2016){Rice}, {Goodman}, {Bergin}, {Beaumont}, \&
  {Dame}}]{2016ApJ...822...52R}
{Rice}, T.~S., {Goodman}, A.~A., {Bergin}, E.~A., {Beaumont}, C., \& {Dame},
  T.~M. 2016, \apj, 822, 52, \dodoi{10.3847/0004-637X/822/1/52}

\bibitem[{{Rodriguez} {et~al.}(1980){Rodriguez}, {Ho}, \&
  {Moran}}]{1980ApJ...240L.149R}
{Rodriguez}, L.~F., {Ho}, P.~T.~P., \& {Moran}, J.~M. 1980, \apjl, 240, L149,
  \dodoi{10.1086/183342}

\bibitem[{{Roman-Duval} {et~al.}(2010){Roman-Duval}, {Jackson}, {Heyer},
  {Rathborne}, \& {Simon}}]{2010ApJ...723..492R}
{Roman-Duval}, J., {Jackson}, J.~M., {Heyer}, M., {Rathborne}, J., \& {Simon},
  R. 2010, \apj, 723, 492, \dodoi{10.1088/0004-637X/723/1/492}

\bibitem[{{Rosolowsky} {et~al.}(2021){Rosolowsky}, {Hughes}, {Leroy}, {Sun},
  {Querejeta}, {Schruba}, {Usero}, {Herrera}, {Liu}, {Pety}, {Saito},
  {Be{\v{s}}li{\'c}}, {Bigiel}, {Blanc}, {Chevance}, {Dale}, {Deger}, {Faesi},
  {Glover}, {Henshaw}, {Klessen}, {Kruijssen}, {Larson}, {Lee}, {Meidt}, {Mok},
  {Schinnerer}, {Thilker}, \& {Williams}}]{2021MNRAS.502.1218R}
{Rosolowsky}, E., {Hughes}, A., {Leroy}, A.~K., {et~al.} 2021, \mnras, 502,
  1218, \dodoi{10.1093/mnras/stab085}

\bibitem[{{Rosolowsky} {et~al.}(2008){Rosolowsky}, {Pineda}, {Kauffmann}, \&
  {Goodman}}]{2008ApJ...679.1338R}
{Rosolowsky}, E.~W., {Pineda}, J.~E., {Kauffmann}, J., \& {Goodman}, A.~A.
  2008, \apj, 679, 1338, \dodoi{10.1086/587685}

\bibitem[{{Sanders} {et~al.}(1985){Sanders}, {Scoville}, \&
  {Solomon}}]{1985ApJ...289..373S}
{Sanders}, D.~B., {Scoville}, N.~Z., \& {Solomon}, P.~M. 1985, \apj, 289, 373,
  \dodoi{10.1086/162897}

\bibitem[{{Schuller} {et~al.}(2017){Schuller}, {Csengeri}, {Urquhart},
  {Duarte-Cabral}, {Barnes}, {Giannetti}, {Hernand ez}, {Leurini}, {Mattern},
  {Medina}, {Agurto}, {Azagra}, {Anderson}, {Beltr{\'a}n}, {Beuther},
  {Bontemps}, {Bronfman}, {Dobbs}, {Dumke}, {Finger}, {Ginsburg}, {Gonzalez},
  {Henning}, {Kauffmann}, {Mac-Auliffe}, {Menten}, {Montenegro-Montes},
  {Moore}, {Muller}, {Parra}, {Perez-Beaupuits}, {Pettitt}, {Russeil},
  {S{\'a}nchez-Monge}, {Schilke}, {Schisano}, {Suri}, {Testi}, {Torstensson},
  {Venegas}, {Wang}, {Wienen}, {Wyrowski}, \& {Zavagno}}]{2017A&A...601A.124S}
{Schuller}, F., {Csengeri}, T., {Urquhart}, J.~S., {et~al.} 2017, \aap, 601,
  A124, \dodoi{10.1051/0004-6361/201628933}

\bibitem[{{Seifried} {et~al.}(2022){Seifried}, {Beuther}, {Walch}, {Syed},
  {Soler}, {Girichidis}, \& {W{\"u}nsch}}]{2022MNRAS.512.4765S}
{Seifried}, D., {Beuther}, H., {Walch}, S., {et~al.} 2022, \mnras, 512, 4765,
  \dodoi{10.1093/mnras/stac607}

\bibitem[{{Shetty} {et~al.}(2010){Shetty}, {Collins}, {Kauffmann}, {Goodman},
  {Rosolowsky}, \& {Norman}}]{2010ApJ...712.1049S}
{Shetty}, R., {Collins}, D.~C., {Kauffmann}, J., {et~al.} 2010, \apj, 712,
  1049, \dodoi{10.1088/0004-637X/712/2/1049}

\bibitem[{{Shimajiri} {et~al.}(2019){Shimajiri}, {Andr{\'e}}, {Palmeirim},
  {Arzoumanian}, {Bracco}, {K{\"o}nyves}, {Ntormousi}, \&
  {Ladjelate}}]{2019A&A...623A..16S}
{Shimajiri}, Y., {Andr{\'e}}, P., {Palmeirim}, P., {et~al.} 2019, \aap, 623,
  A16, \dodoi{10.1051/0004-6361/201834399}

\bibitem[{{Snell} {et~al.}(1980){Snell}, {Loren}, \&
  {Plambeck}}]{1980ApJ...239L..17S}
{Snell}, R.~L., {Loren}, R.~B., \& {Plambeck}, R.~L. 1980, \apjl, 239, L17,
  \dodoi{10.1086/183283}

\bibitem[{{Sokolov} {et~al.}(2020){Sokolov}, {Pineda}, {Buchner}, \&
  {Caselli}}]{2020ApJ...892L..32S}
{Sokolov}, V., {Pineda}, J.~E., {Buchner}, J., \& {Caselli}, P. 2020, \apjl,
  892, L32, \dodoi{10.3847/2041-8213/ab8018}

\bibitem[{{Solomon} {et~al.}(1987){Solomon}, {Rivolo}, {Barrett}, \&
  {Yahil}}]{1987ApJ...319..730S}
{Solomon}, P.~M., {Rivolo}, A.~R., {Barrett}, J., \& {Yahil}, A. 1987, \apj,
  319, 730, \dodoi{10.1086/165493}

\bibitem[{{Solomon} {et~al.}(1979){Solomon}, {Sanders}, \&
  {Scoville}}]{1979IAUS...84...35S}
{Solomon}, P.~M., {Sanders}, D.~B., \& {Scoville}, N.~Z. 1979, in IAU
  Symposium, Vol.~84, The Large-Scale Characteristics of the Galaxy, ed. W.~B.
  {Burton}, 35

\bibitem[{{Solomon} \& {Vanden Bout}(2005)}]{2005ARA&A..43..677S}
{Solomon}, P.~M., \& {Vanden Bout}, P.~A. 2005, \araa, 43, 677,
  \dodoi{10.1146/annurev.astro.43.051804.102221}

\bibitem[{{Su} {et~al.}(2019){Su}, {Yang}, {Zhang}, {Gong}, {Wang}, {Zhou},
  {Wang}, {Chen}, {Sun}, {Chen}, {Xu}, \& {Jiang}}]{2019ApJS..240....9S}
{Su}, Y., {Yang}, J., {Zhang}, S., {et~al.} 2019, \apjs, 240, 9,
  \dodoi{10.3847/1538-4365/aaf1c8}

\bibitem[{{Sun} {et~al.}(2015){Sun}, {Xu}, {Yang}, {Li}, {Du}, {Zhang}, \&
  {Zhou}}]{2015ApJ...798L..27S}
{Sun}, Y., {Xu}, Y., {Yang}, J., {et~al.} 2015, \apjl, 798, L27,
  \dodoi{10.1088/2041-8205/798/2/L27}

\bibitem[{{Sun} {et~al.}(2021){Sun}, {Yang}, {Yan}, {Lin}, {Zhang}, {Su}, {Xu},
  {Chen}, {Wang}, \& {Zhou}}]{2021ApJS..256...32S}
{Sun}, Y., {Yang}, J., {Yan}, Q.-Z., {et~al.} 2021, \apjs, 256, 32,
  \dodoi{10.3847/1538-4365/ac11fe}

\bibitem[{{Tan} {et~al.}(2018){Tan}, {Gao}, {Zhang}, {Greve}, {Jiang},
  {Wilson}, {Yang}, {Bemis}, {Chung}, {Matsushita}, {Shi}, {Ao}, {Brinks},
  {Currie}, {Davis}, {de Grijs}, {Ho}, {Imanishi}, {Kohno}, {Lee}, {Parsons},
  {Rawlings}, {Rigopoulou}, {Rosolowsky}, {Bulger}, {Chen}, {Chapman}, {Eden},
  {Gear}, {Gu}, {He}, {Jiao}, {Liu}, {Liu}, {Li}, {Micha{\l}owski},
  {Nguyen-Luong}, {Qiu}, {Smith}, {Violino}, {Wang}, {Wang}, {Wang}, {Yeh},
  {Zhao}, \& {Zhu}}]{2018ApJ...860..165T}
{Tan}, Q.-H., {Gao}, Y., {Zhang}, Z.-Y., {et~al.} 2018, \apj, 860, 165,
  \dodoi{10.3847/1538-4357/aac512}

\bibitem[{{Traficante} {et~al.}(2018){Traficante}, {Duarte-Cabral}, {Elia},
  {Fuller}, {Merello}, {Molinari}, {Peretto}, {Schisano}, \& {Di
  Giorgio}}]{2018MNRAS.477.2220T}
{Traficante}, A., {Duarte-Cabral}, A., {Elia}, D., {et~al.} 2018, \mnras, 477,
  2220, \dodoi{10.1093/mnras/sty798}

\bibitem[{{Ungerechts} \& {Thaddeus}(1987)}]{1987ApJS...63..645U}
{Ungerechts}, H., \& {Thaddeus}, P. 1987, \apjs, 63, 645,
  \dodoi{10.1086/191176}

\bibitem[{Virtanen {et~al.}(2020)Virtanen, Gommers, Oliphant, Haberland, Reddy,
  Cournapeau, Burovski, Peterson, Weckesser, Bright, {van der Walt}, Brett,
  Wilson, Millman, Mayorov, Nelson, Jones, Kern, Larson, Carey, Polat, Feng,
  Moore, {VanderPlas}, Laxalde, Perktold, Cimrman, Henriksen, Quintero, Harris,
  Archibald, Ribeiro, Pedregosa, {van Mulbregt}, \& {SciPy 1.0
  Contributors}}]{2020SciPy-NMeth}
Virtanen, P., Gommers, R., Oliphant, T.~E., {et~al.} 2020, Nature Methods, 17,
  261, \dodoi{10.1038/s41592-019-0686-2}

\bibitem[{{Wang} {et~al.}(2019){Wang}, {Yang}, {Su}, {Du}, {Ma}, \&
  {Zhang}}]{2019ApJS..243...25W}
{Wang}, C., {Yang}, J., {Su}, Y., {et~al.} 2019, \apjs, 243, 25,
  \dodoi{10.3847/1538-4365/ab2d2e}

\bibitem[{{Westmeier} {et~al.}(2012){Westmeier}, {Popping}, \&
  {Serra}}]{2012PASA...29..276W}
{Westmeier}, T., {Popping}, A., \& {Serra}, P. 2012, \pasa, 29, 276,
  \dodoi{10.1071/AS11041}

\bibitem[{{Williams} {et~al.}(1994){Williams}, {de Geus}, \&
  {Blitz}}]{1994ApJ...428..693W}
{Williams}, J.~P., {de Geus}, E.~J., \& {Blitz}, L. 1994, \apj, 428, 693,
  \dodoi{10.1086/174279}

\bibitem[{{Wilson} {et~al.}(2005){Wilson}, {Dame}, {Masheder}, \&
  {Thaddeus}}]{2005A&A...430..523W}
{Wilson}, B.~A., {Dame}, T.~M., {Masheder}, M.~R.~W., \& {Thaddeus}, P. 2005,
  \aap, 430, 523, \dodoi{10.1051/0004-6361:20035943}

\bibitem[{{Wilson} {et~al.}(1970){Wilson}, {Jefferts}, \&
  {Penzias}}]{1970ApJ...161L..43W}
{Wilson}, R.~W., {Jefferts}, K.~B., \& {Penzias}, A.~A. 1970, \apjl, 161, L43,
  \dodoi{10.1086/180567}

\bibitem[{{Wong} \& {Blitz}(2002)}]{2002ApJ...569..157W}
{Wong}, T., \& {Blitz}, L. 2002, \apj, 569, 157, \dodoi{10.1086/339287}

\bibitem[{{Xu} {et~al.}(2006){Xu}, {Reid}, {Zheng}, \&
  {Menten}}]{2006Sci...311...54X}
{Xu}, Y., {Reid}, M.~J., {Zheng}, X.~W., \& {Menten}, K.~M. 2006, Science, 311,
  54, \dodoi{10.1126/science.1120914}

\bibitem[{{Yan} {et~al.}(2020){Yan}, {Yang}, {Su}, {Sun}, \&
  {Wang}}]{2020ApJ...898...80Y}
{Yan}, Q.-Z., {Yang}, J., {Su}, Y., {Sun}, Y., \& {Wang}, C. 2020, \apj, 898,
  80, \dodoi{10.3847/1538-4357/ab9f9c}

\bibitem[{{Yan} {et~al.}(2021{\natexlab{a}}){Yan}, {Yang}, {Su}, {Sun}, {Xu},
  {Wang}, {Zhou}, \& {Wang}}]{2021ApJ...922....8Y}
{Yan}, Q.-Z., {Yang}, J., {Su}, Y., {et~al.} 2021{\natexlab{a}}, \apj, 922, 8,
  \dodoi{10.3847/1538-4357/ac214f}

\bibitem[{{Yan} {et~al.}(2021{\natexlab{b}}){Yan}, {Yang}, {Yang}, {Sun}, \&
  {Wang}}]{2021ApJ...910..109Y}
{Yan}, Q.-Z., {Yang}, J., {Yang}, S., {Sun}, Y., \& {Wang}, C.
  2021{\natexlab{b}}, \apj, 910, 109, \dodoi{10.3847/1538-4357/abe628}

\bibitem[{{Yang} {et~al.}(2018){Yang}, {Thompson}, {Urquhart}, \&
  {Tian}}]{2018ApJS..235....3Y}
{Yang}, A.~Y., {Thompson}, M.~A., {Urquhart}, J.~S., \& {Tian}, W.~W. 2018,
  \apjs, 235, 3, \dodoi{10.3847/1538-4365/aaa297}

\bibitem[{{Yang} {et~al.}(2022){Yang}, {Urquhart}, {Wyrowski}, {Thompson},
  {K{\"o}nig}, {Colombo}, {Menten}, {Duarte-Cabral}, {Schuller}, {Csengeri},
  {Eden}, {Barnes}, {Traficante}, {Bronfman}, {Sanchez-Monge}, {Ginsburg},
  {Cesaroni}, {Lee}, {Beuther}, {Medina}, {Mazumdar}, \&
  {Henning}}]{2022A&A...658A.160Y}
{Yang}, A.~Y., {Urquhart}, J.~S., {Wyrowski}, F., {et~al.} 2022, \aap, 658,
  A160, \dodoi{10.1051/0004-6361/202142039}

\bibitem[{{Yang} \& {Fukui}(1992)}]{1992ApJ...386..618Y}
{Yang}, J., \& {Fukui}, Y. 1992, \apj, 386, 618, \dodoi{10.1086/171043}

\bibitem[{{Zhang} {et~al.}(2001){Zhang}, {Hunter}, {Brand}, {Sridharan},
  {Molinari}, {Kramer}, \& {Cesaroni}}]{2001ApJ...552L.167Z}
{Zhang}, Q., {Hunter}, T.~R., {Brand}, J., {et~al.} 2001, \apjl, 552, L167,
  \dodoi{10.1086/320345}

\bibitem[{{Zhang} {et~al.}(2020){Zhang}, {Yang}, {Xu}, {Chen}, {Su}, {Sun},
  {Zhou}, {Li}, \& {Lu}}]{2020ApJS..248...15Z}
{Zhang}, S., {Yang}, J., {Xu}, Y., {et~al.} 2020, \apjs, 248, 15,
  \dodoi{10.3847/1538-4365/ab879a}

\end{thebibliography}





%

\end{document}